\documentclass{jaa}
\usepackage{graphicx}
\usepackage{caption}
\usepackage[font=scriptsize]{caption}
\usepackage{subcaption}
\usepackage{amssymb}
\usepackage{natbib}
\usepackage{array}
\usepackage{float}
\usepackage{url}
\usepackage[colorlinks=true,citecolor=blue,urlcolor=blue,linkcolor=blue]{hyperref}

\begin{document}\sloppy

\title{Photometric and Kinematic study of the open cluster NGC~1027}
\author{Apara Tripathi\textsuperscript{1}, Neelam Panwar\textsuperscript{2}, Saurabh Sharma\textsuperscript{2}, Brijesh Kumar\textsuperscript{2}, Shantanu Rastogi\textsuperscript{1}}
\affilOne{\textsuperscript{1}D D U Gorakhpur University, Gorakhpur, Uttar Pradesh,India\\}
\affilTwo{\textsuperscript{2}Aryabhatta Research Institute of Observational Sciences, Nainital, Uttarakhand}

\twocolumn[{
\maketitle

\corres{aparatrip@gmail.com}
\msinfo{2022}{}

\begin{abstract}
 We present photometric and kinematic analysis of an intermediate age open cluster NGC 1027 using $UBV(RI)_c$ and Gaia Early Data Release 3 (EDR3) data. Structural and fundamental parameters such as cluster center, cluster extent, reddening, age and distance are estimated in this study. Cluster center is found about 2 arcmin away from the center reported earlier. Radius has been estimated to be about 8.00 arcmin(2.65pc). Using proper motion Gaia EDR3 data, membership probabilities has been derived for the stars in the region of cluster radius.  We find mean proper motion of the cluster to be $\sim$(-0.84, 2.04) mas yr$^{-1}$ in (RA, DEC). We find 217 most probable (P$_\mu>$ 70\%) cluster members with mean parallax 0.892 $\pm$ 0.088 mas. Out of these, 160 members have counterparts in our optical observations. Few stars having P$_\mu>$ 70\% are found out of the cluster radius showing imprints of dynamical evolution. The color-color and color-magnitude diagrams for the cluster members found within 8.00 arcmin have been constructed using $UBV(RI)_c$ photometry and Gaia EDR3 data. This yields a reddening E($B$-$V$) $\sim$ 0.36 mag, age $\sim$ 130 Myr and distance $\sim$ 1.14 kpc. The mass function slope in the cluster region is $\Gamma$ $\sim$ -1.46 $\pm$ 0.15, which is similar to other Galactic open clusters. The dynamical study shows lack of faint stars in its inner region leading to mass segregation effect. A comparison of dynamical age with cluster age indicates that NGC 1027 is a dynamically relaxed cluster suggesting that mass segregation may be imprint of its dynamical relaxation.
 
\end{abstract}
}]

\doinum{}
\artcitid{\#\#\#\#}
\volnum{000}
\year{2022}
\pgrange{1--}
\setcounter{page}{1}
\lp{1}

\section{Introduction}
 Stars in the Galaxy often form in star clusters. They contain stars of diverse mass range that are gravitationally bound and presumably formed from the same molecular material, at same time and located at the same distance and move together \citep{Harris1994,Friel1995,Bate2003,Carraro2007a,CantatGaudin2020,Pang2020}. Observations suggest that the age of Galactic star clusters range from a few Myr to few Gyr, indicating that they have been forming all along the life of our Galaxy. External forces of interaction with the galactic tidal field and two body relaxation effect changes the structure of open cluster probing them to be good targets to investigate astrophysical processes involved in formation and evolution of stars and Galaxies \citep{Miholics2014,apara2013,apara2014}.
 
Intermediate-age open clusters (OCs) are located in the disc of the Galaxy. Study of these type of OCs play a crucial role in advancing our understanding of the physical processes involved such as change in its structure, mass function, mass segregation and develop the mechanism that govern their evolution by time. They probe the dynamical and chemical evolution of the Galactic disc in age and metallicity. Their study also improvise important constraints for the evolutionary astrophysical models.
\citep{Sagar1986,Phelps1994,Lada&Lada2003,Chen2003,Joshi2016,CastroGinard2018,Angelo2019,CantatGaudin2020}

 NGC~1027  ($\alpha_{J2000}$ - 02:42:40,  $\delta_{J2000}$ +61:35:42; l $\sim$ 135.758 and b $\sim$ + 1.525 ), also known as Melotte~16, Collinder 30, OCL 357, MWSC 225 open star cluster, is located in the eastern part of Cassiopeia constellation near the northern celestial pole,  adjacent to two prominent nebulae, the Heart (IC 1805) and Soul Nebulae (IC 1848). However, NGC~1027 is not physically associated with these two nebulae as they are more distant \citep[d $\sim$ 2.1 kpc;][]{Chauhan2011,Panwar2017} compared to the NGC~1027 \citep[d$\sim$ 1 kpc;][]{Maciejewski2007}.
Some basic parameters of the cluster as reported in WEBDA, A site Devoted to Stellar Clusters in the Galaxy and the Magellanic Clouds are listed in Table \ref{tab1}.\citet{Ruprecht1966} considered this cluster as class {\sc iii} 2 p, i.e., a less populated open cluster with noticeable concentration and a medium range in magnitude of the stars \citet{Trumpler1930}.

Based on the wide-field $BV$ CCD photometric survey, \citet{Maciejewski2007} reported basic parameters of 42 open clusters (OCs) including NGC 1027. They estimated reddening E($B$-$V$) $\sim$ 0.41$\pm${0.12} mag, log age $\sim$ 8.4 and distance $\sim$ 1.03 kpc for NGC 1027. \citet{Liu2019} cataloged 2443 cluster candidates using Star cluster Hunting Pipeline (SHiP) and classified NGC~1027 as an old class {\sc iii} star cluster with a prominent MS turn-off in the color-magnitude diagram (CMD). He suggested for re investigation of Class~{\sc iii} star cluster candidates. 

Photometric studies of OCs are found to be fundamental tools for determining their structural parameters. These parameters play an important role towards understanding galactic structure and evolution. However, most of the OCs are affected by field star contamination. Therefore finding out member stars of the cluster becomes necessary to look into the cluster properties \citep{Carraro2008,Kharchenko2013,CantatGaudin2018}. Several methods by different researches has been suggested to determine membership of cluster stars \citep{MedhiandTamura2013,Gao2014}. Proper motion diagram is a reliable tool to identify cluster members on the basis of astrometric precision and membership determination on kinematic method \citep[e.g.,][]{CantatGaudin2018,CastroGinard2018,CantatGaudin2019,Bisht2019,CastroGinard2020,Joshi2020}. This precise learning of cluster membership gives information about distribution of stellar mass in the OCs and throw light on dynamic evolution of star clusters. Numerous research studies have obtained present-day MF for a number of OCs \citep{Dib2017,Bostanc2018,Joshi2020}, yet, it is still not clear weather initial mass function (IMF) is universal in time and space or depends upon different star forming conditions \citep{Elmegreen2000,Kroupa2002,Bastian2010}. Study of the distribution of low and high mass stars towards the cluster region provides a clue to the mass segregation in OCs.

In the present work $UBVR_cI_c$ photometry for NGC 1027 has been done for the first time. We investigate the spatial structure, basic parameters, extent, reddening and age to understand the dynamical evolution of the cluster using the kinematic Gaia EDR3 data. Our high quality CCD photometry data go down to $V$ $\sim$ 22 mag. We estimate membership probabilities of stars in the cluster NGC 1027 down to $G \sim$ 20 mag. The high probability members are utilized to examine mass function and mass segregation in the cluster NGC $\sim$ 1027.

   \begin{table}[ht]
   \centering
   \caption{Parameters listed in WEBDA for cluster NGC 1027.}
   \begin{tabular}[t]{llll}
   \hline
     Cluster Parameter& Values\\
     \hline
     Trumpler class & IIIp\\
     RA(J2000)& 05:36:18\\
     DEC(J2000)& +34:08:24\\
     Longitude(l/deg)&164.535\\
     Latitude(b/deg)&1.072\\
     Distance(d/pc)&1318\\
     Reddening(E(B-V)/mag)&0.36\\
     Age(Log(t/yr))&7.487\\
\hline 
\end{tabular}
\label{tab1}
\end{table}

The organization of the manuscript is as follows. Section 2 provides a brief description of the observations, data reduction procedures and astrometric calibration. Results and analysis including radial density profile, kinematic members, reddening, CMDs, age and distance to the cluster is discussed in Section 3. We discuss the results including a discussion on the presence of kinematic members beyond the cluster radius in section 4 and Section 5 summarises the results.

\section{Observations \& Data Reduction}
\subsection{UBVR\textsubscript{c}I\textsubscript{c} photometric observations}
The CCD UBVR\textsubscript{c}I\textsubscript{c} data used in this study is taken using 104-cm Sampurnanand Telescope (ST) of the Aryabhatta Research Institute of Observational Sciences (ARIES), Nainital. The telescope covers about 13 $\times$ 13 arcmin$^2$ on the sky. In order to improve signal-to-noise ratio, binned mode of 2 $\times$ 2 pixel is used during the observations. The observing log is given in Table \ref{tab2}. Several number of bias and twilight flat-field frames were taken in all the filters along with multiple short and long exposure frames of the target and standard fields. To calibrate the magnitude of stars in cluster field standard field SA~104 of \citet{Landolt1992} is observed.  Basic steps are  performed using the standard tasks within the IRAF data reduction package for pre-processing of data frames which includes bias subtraction, flat fielding and  cosmic ray removal.

\begin{table}
\caption{Description of the optical observation for NGC 1027 and standard field, with dates and exposure times for each passband}
   \small
   \begin{tabular}{ccccccc}
   \hline
     Cluster/&Date&Filter&Exp(sec)\\Standard field
     &&&x no.of Frames\\
     \hline
     NGC 1027&24/25 Jan 06&V&120x3\\
                       &&B&180x3\\
                       &&I&60x2\\
                       &&R&60x2,300x3\\
                       &&U&300x2\\
            &24 Dec 05&V&900x2\\
                       &&B&1200x3\\
                       &&I&240x3\\
                       &&R&300x3\\
                       &&U&1800x2\\

   SA104&24/25 Jan 06& V&130x5, 180x1\\
                       &&B&200x5,240x1\\
                       &&I&60x7\\
                       &&R&100x5,120x2\\
                       &&U&300x4, 240x1\\
\hline 
\end{tabular}
\label{tab2}
\end{table}

Airmass covers range of 1.2 to 1.3 for \citet{Landolt1992} standards. 9 standard stars of \citet{Landolt1992} is taken from the observed field of SA~104. The mean FWHM varies about 2 arcsec over the entire obsation period.

\subsection{Photometric and Astrometric Calibration}
Photometry of bias-subtracted and flat field corrected CCD frames is carried out using DAOPHOT-{\sc ii} software\citep{Stetson1987,Stetson2000}. Quantitative values for the brightness of stars are obtained by aperture photometry and profile fitting photometry. 

Measurements of bright stars which are saturated in deep exposure frames have been taken from short-exposure frames. For translating the observed aperture magnitudes to the standard magnitudes, least-square linear regressions outlined by \cite{Stetson2000}, are fitted. 
The instrumental magnitude calibration equations are as follows:\\
 $ v= V + 4.22\pm 0.01-(0.04\pm 0.01)(B-V)+(0.21\pm 0.01)X\\$
 $ b= B + 4.65\pm 0.01-(0.06\pm 0.01)(B-V)+(0.33\pm 0.01)X\\$
 $ i= I_c + 4.67\pm 0.01-(0.02\pm 0.01)(V-I_c)+(0.09\pm 0.01)X\\$
 $ r= R_c + 4.13\pm 0.01-(0.04\pm 0.01)(V-R_c)+(0.16\pm 0.01)X\\$
 $ u= U + 6.86\pm 0.01-(0.03\pm 0.01)(U-B)+(0.61\pm 0.01)X\\$
here V, B, $R_c,I_c$ and U are standard magnitudes and v, b, i, r, u represent instrumental aperture magnitudes normalized for 1 sec of exposure time and X is the airmass. Second order color correction terms are ignored as they are small in comparision to the photometric data reduction error. Second order color correction terms are  small in comparison to the other photometric data reduction error. Hence, they are ignored. Zero-points for local standards are evaluated accounting the aperture growth curve, difference in exposure times and atmospheric extinction. Errors in zero-points and color coefficients are $\sim$
0.01 mag.  A final catalogue of stellar objects identified in atleast two filters were obtained with shape-defining parameters 0.2 $\leq$ sharpness 
$<$ 1.0 and goodness of fit estimator $\chi$ $<$ 5. The accuracy of the morphology and position of the cluster features in CMDs depends on the variation of photometric errors with magnitudes. Photometric errors are approximated to be related with the signal-to-noise ratio of the source. 

\begin{figure*}
\includegraphics[scale = 0.5, trim = 0 0 0 0, clip]{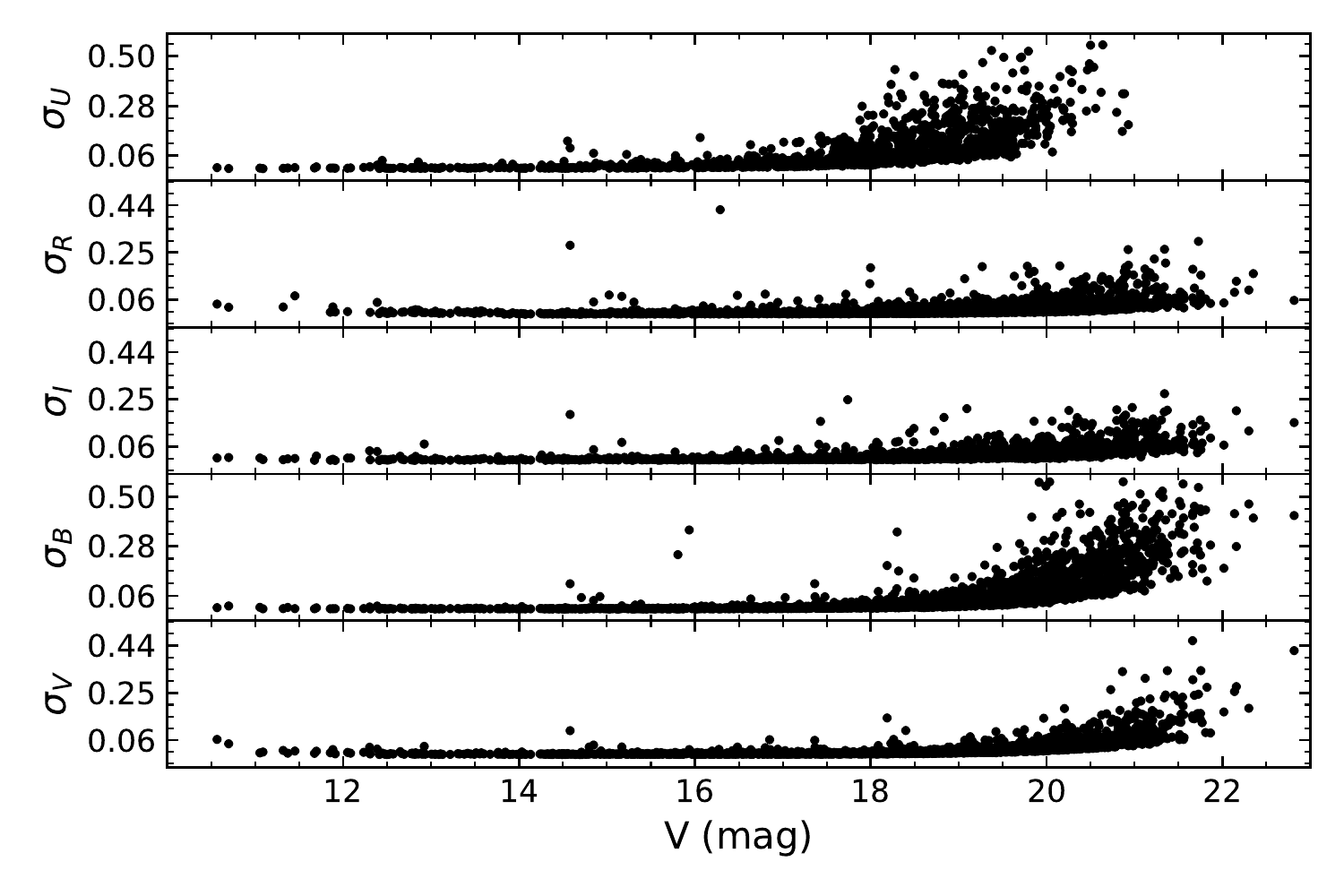}
\includegraphics[scale = 0.5, trim = 0 0 0 0, clip]{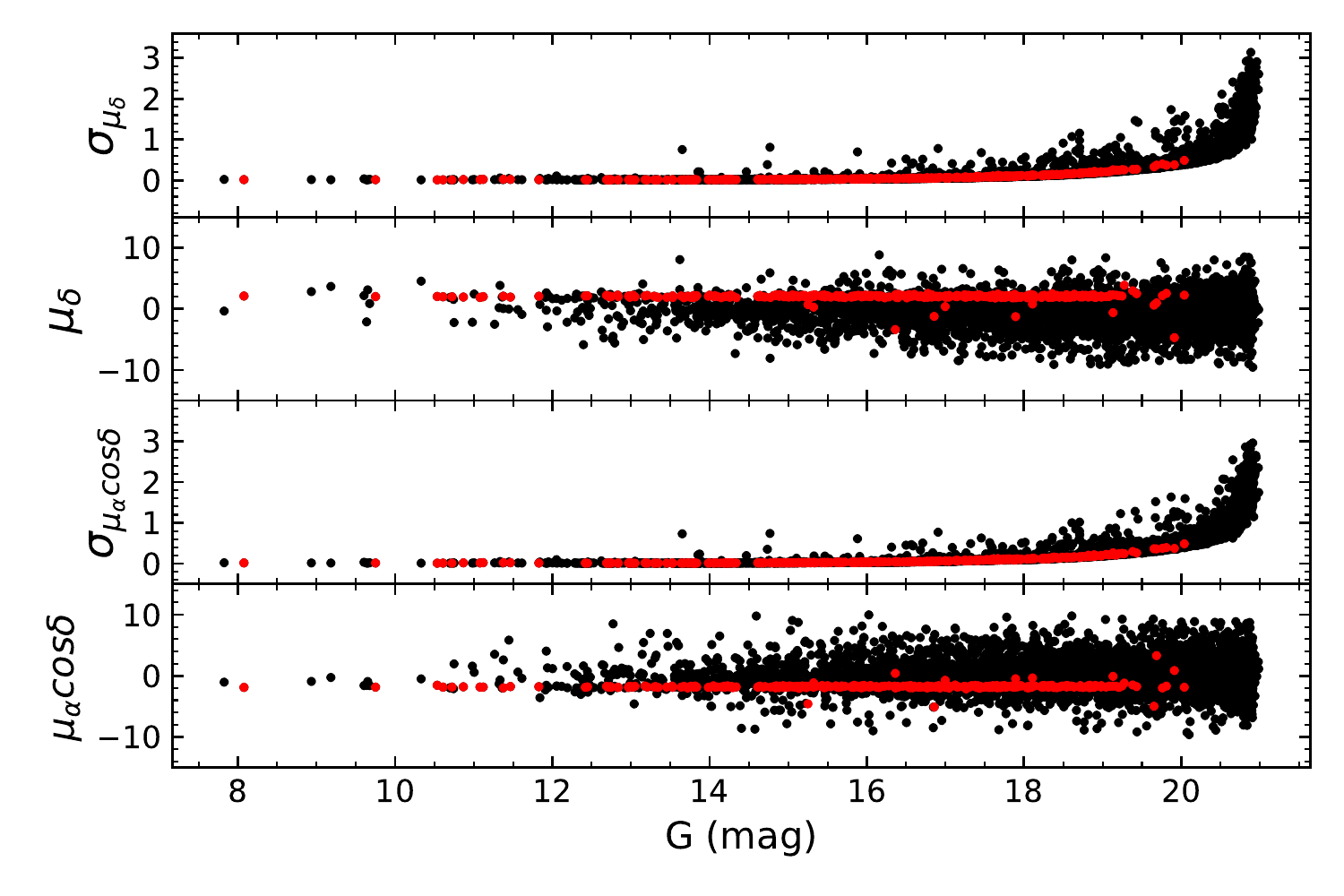}
\caption{Left panel shows photometric errors corresponding
to the brightness measurement at $U$, $B$, $V$, $R_c$, and
$I_c$ against $V$-band brightness.
Errors on the y-axis represent the internal error as
estimated by $DAOPHOT$ routine. Right panel shows the
plot of proper motions in both $RA$ and $Declination$ directions and their
errors versus $G$ magnitude. Blue symbol represent stars having membership 
probability $>$ 70$\%$.}
\label{fig1}
\end{figure*}

The internal errors as derived from $DAOPHOT$ in $U$, $B$, $V$, $R_c$ and $I_c$ are plotted against $V$ mag in Figure \ref{fig1}. This shows that the photometric error is less than 0.01 mag in $B$ up to $V$ $\sim$17 mag; less than 0.01 mag in $V$, $R$ and $I$ up to $V$ $\sim$18 mag; and for $U$-band it is less than 0.01 mag up to $V$ $\sim$16 mag. \\

Astrometric solutions are used to get celestial coordinates of all the stars in J2000.0. 
 We use the $CCMAP$ and $CCTRAN$ tasks of $IRAF$ to obtain the celestial coordinates of the detected stars. The sources detected in different wavebands are  band merged using a matching radius of $\sim$ 2 arcsec. 
In total, 5913 stars are detected in all five bands. 
To determine the completeness of our $UBVR_CI_C$ data, we used the magnitude histograms \citep{Panwar2017}. We found that completeness of the data in $V$, $B$, $I_c$, $R_C$ and $U$ band is limited to about 20 mag, 20.5 mag, 18 mag, 19 mag and  20 mag, respectively.

\subsection{Kinematic data from Gaia EDR3}
The Early Gaia Data Release 3 \citep{EDR3} came to public in the year 2020. This data consists of five parametric astrometric solutions which include positions on the sky ($\alpha$, $\delta$) parallaxes and proper motion (PM) ($\mu_\alpha$cos $\delta$, $\mu\delta$) with a limiting magnitude of G $\sim$21 mag. The Gaia EDR3 data are comparatively more accurate than the Second Gaia Data Release \citep{GDR2}.

For analysis we have cross matched optical CCD data with Gaia EDR3 \citep{EDR3}. Parallax uncertainties are in the range up to 0.02-0.03 milliarcsec (mas) for sources at G $\le$ 15 mag and 0.07 mas for sources at G $\sim$ 17 mag. The uncertainties in the respective proper motion components are upto 0.01 - 0.02 mas $yr^{-1}$ for G  $\le$ 15 mag, 0.05 mas $yr^{-1}$ for $\sim$ 20 mag. Proper motion and their corresponding errors are plotted against Gmag in right panel of Figure \ref{fig1}.  

\section{Data Analysis \& Results}
\subsection {Cluster Center and Radial Density Profile}

To obtain the cluster parameters, first we estimate the cluster center using the $Python$ Kernel density estimation (KDE). For this, we use the Gaia EDR3 data for the sources with G-band magnitude uncertainty $\le$0.1 mag within 40 arcmin radius around the cluster center as mentioned in the $WEBDA$. Our analysis shows the cluster center at $\alpha$, $\delta$ $\sim$ 02:42:43.9, +61:39:34 (about 2 arcmin away from the center mentioned in $WEBDA$). Cluster radius ($r_{cl}$) is taken as the distance from cluster center to where the average cluster contribution becomes negligible with respect to the background stellar field.

Cluster radius as well as extent of field-star contamination is estimated using spatial surface density profile of stars. For quantitative estimation of the cluster size we have taken the concentric circular regions around the estimated cluster center. Star counts within each annular ring of width  about 36 arcsec is divided by the respective areas of the annular rings to obtain the number density. To estimate the structural parameters, \citet{King1962} surface density profile is fitted to the radial distribution of stars. The fit is done using a non-linear least squares fit routine which uses the errors as weight. The radial density profile can be represented by :

\begin{equation}
f(r) = \frac{f_0}{1+(\frac{r}{r_c})^2} + f_b
\end{equation}
where f$_0$ is the central density region, r$_c$ is the core radius of the cluster and
f$_b$ is the background density.

\begin{figure*}
	\includegraphics[scale = 0.48, trim = 0 0 0 0, clip]{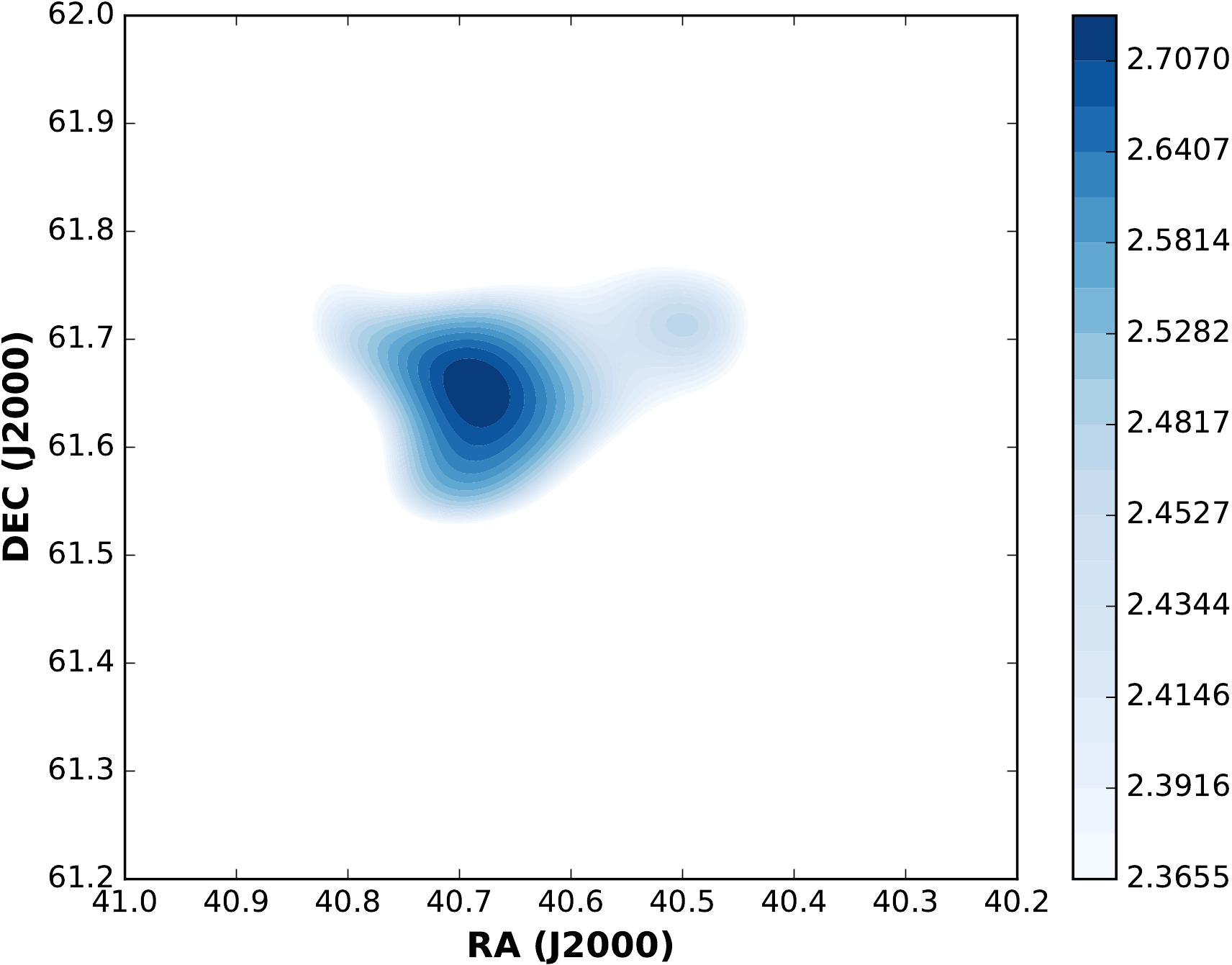}
	\includegraphics[scale = 0.48, trim = 0 0 0 0, clip]{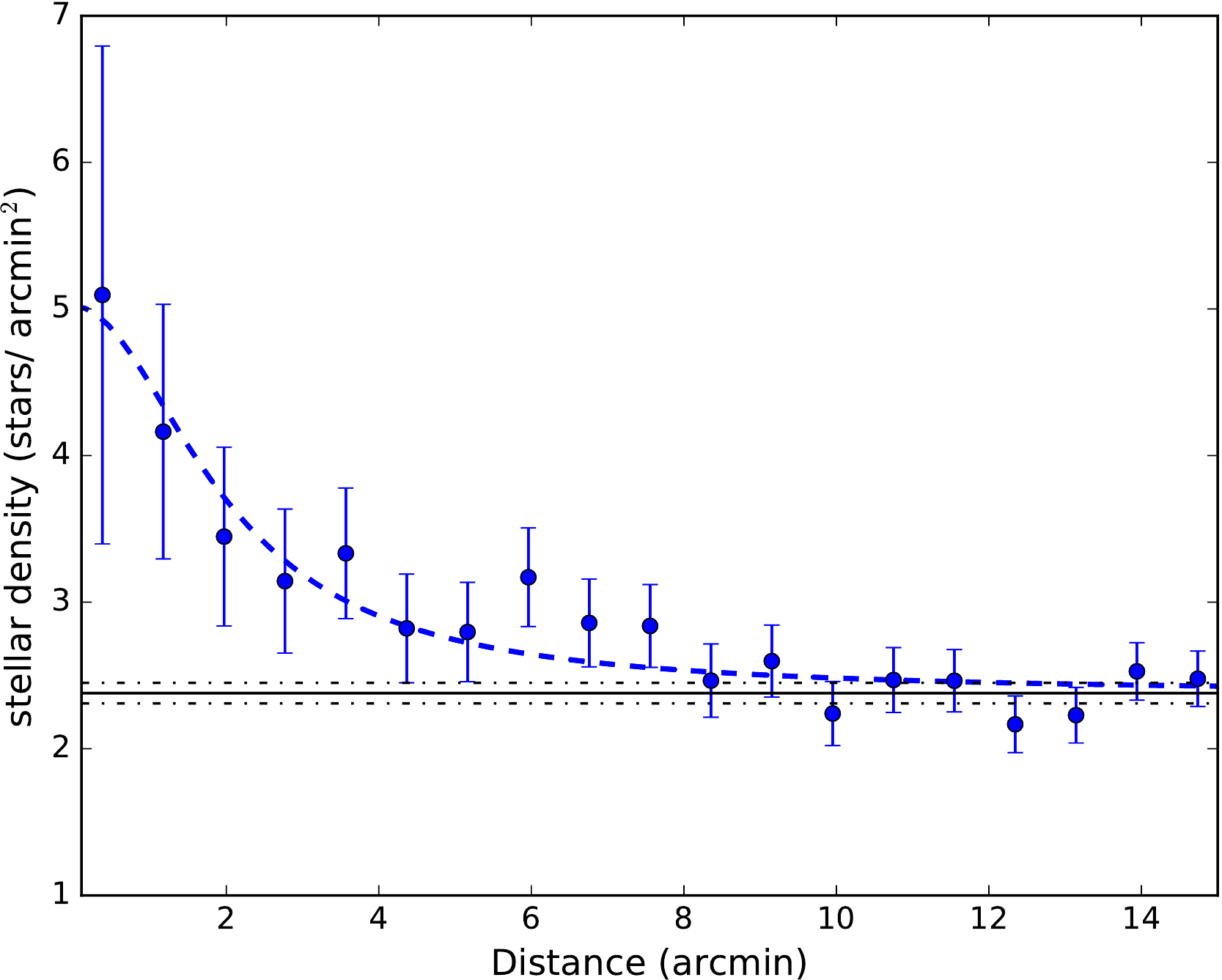}
	\caption{Left panel: The color maps are two-dimensional kernel density estimations (KDEs) of the spatial distribution of the bright GAIA EDR3 sources (G $\le$ 17 mag). Right panel: The radial density profile for the bright stars in the region. Dashed curve represents the least-square fit of the
	\citet{King1962} profile to the observed data points. The error bars represent $\pm$ $\sqrt{N}$ errors. The black line indicates the density of field stars.}
    \label{fig2}
\end{figure*}

The best fit solution along with uncertainties for density
distribution is shown in Figure \ref{fig2}. The radial density profile of the cluster NGC 1027 decreases and flattens around 8 arcmin. Therefore, for further analyses we have considered the cluster radius as 8 arcmin, which is lower than 10.3 arcmin, the value reported before by \citet{Maciejewski2007}. 
For NGC 1027, f$_0$ = 2.2 stars/arcmin$^{-2}$, r$_c$ = 2.4 arcmin and f$_b$ = 2.3 stars/arcmin$^{-2}$.

\subsection{Kinematic members of the cluster NGC~1027 with Gaia EDR3}
\begin{figure*}
\centering
\includegraphics[scale = 0.5, trim = 0 0 0 0, clip]{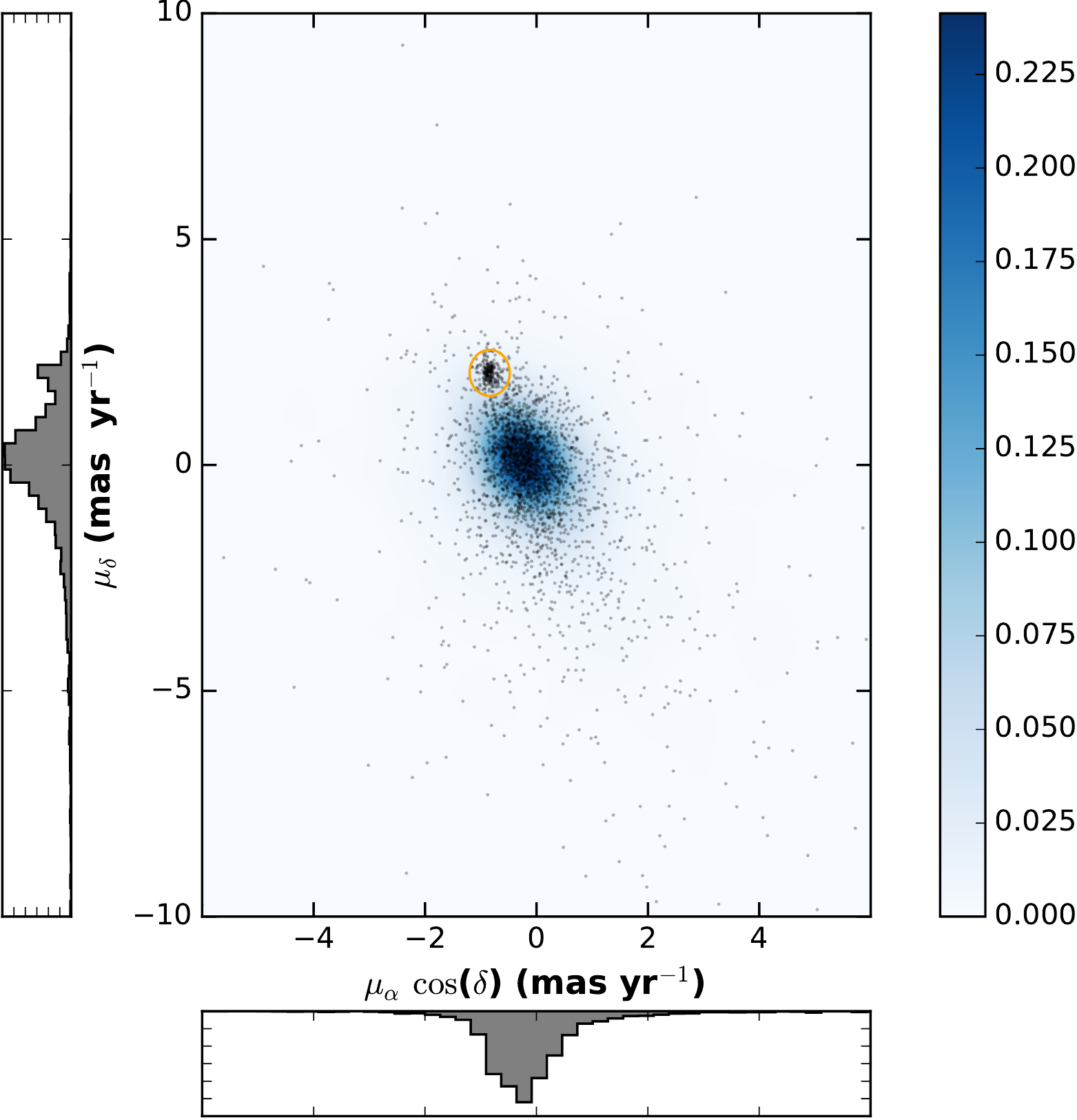}
	\caption{Proper motion vector-point diagram for the Gaia sources located within the cluster radius. Blue color bar shows the distribution of the proper motions for the stars in the control field of the similar area.}
\label{fig3}
\end{figure*}

We use Gaia EDR3 data for the sources within the cluster region  to select the cluster members and to determine mean proper motion of cluster NGC 1027 (radius $\le$ 8 arcmin). We use the proper motion in RA ($\mu_\alpha^\star$) and proper motion in declination ($\mu$ $_\delta$) for the stars within the cluster radius  to generate the vector point diagram (VPD), where $\mu_\alpha^\star$ $\equiv$ ${\mu}_{\alpha}\cos(\delta)$.
We consider only those sources with proper motion uncertainty less than 0.5 mas yr$^{-1}$ and G-band magnitude uncertainty less than 0.1 mag. 
The VPD for those sources is shown in Figure \ref{fig3}. The over-density of sources can be easily noticed in Figure \ref{fig3}.  The proper motion of the stars in the cluster region 
peaks at $(\mu_\alpha^\star$, $\mu_\delta$) $\sim$(-0.84, 2.04) mas yr$^{-1}$. 

\begin{figure}
\includegraphics[scale = 0.48, trim = 0 0 0 0, clip]{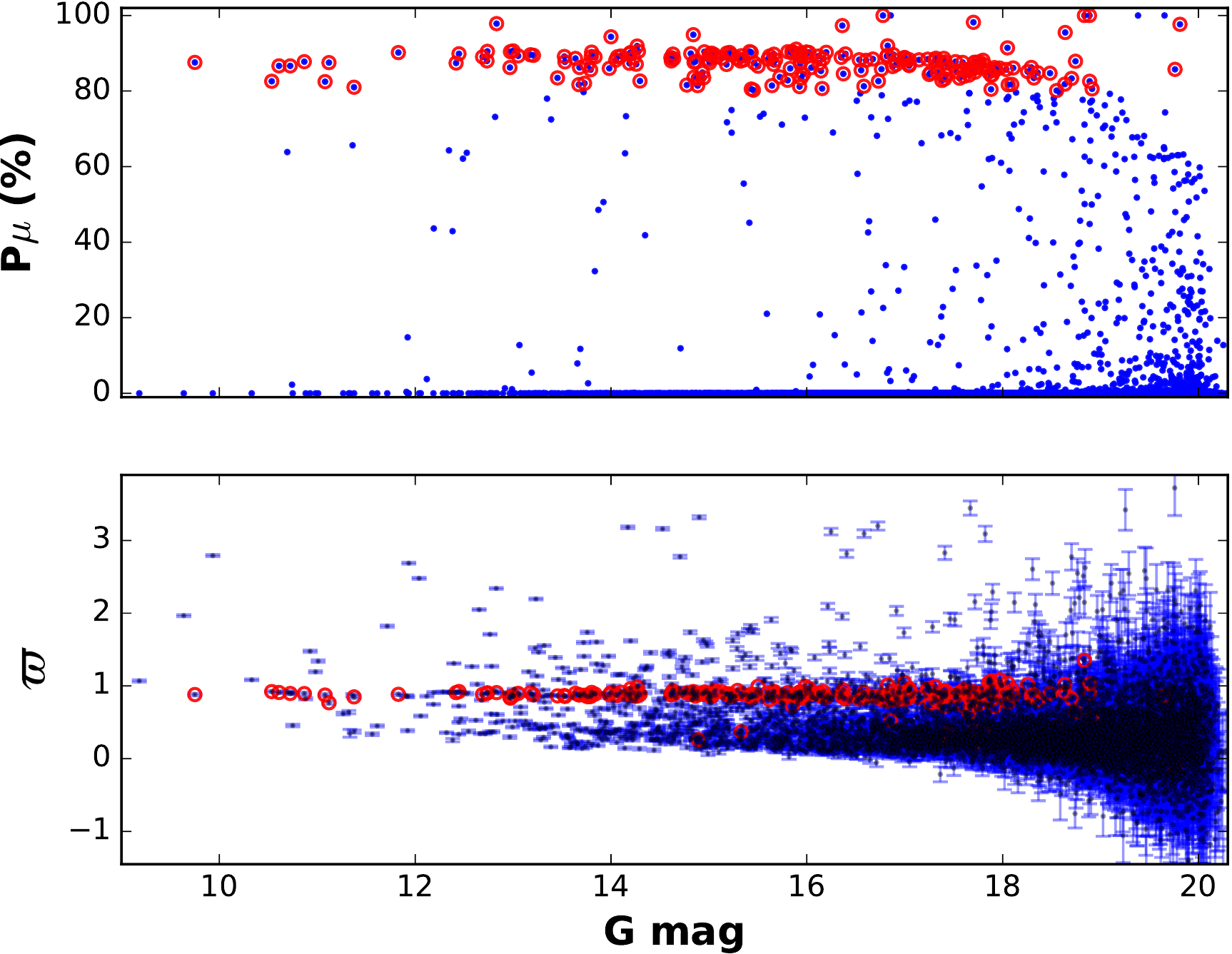}
	\caption{Top Panel: Membership probability for the stars located within the cluster area (radius $\sim$ 8$^{\prime}$) plotted as a function 
	of G-band magnitude. Bottom panel: The blue dots show the distribution of 
	parallax values for the Gaia sources within the cluster as a function of G magnitude. The error bars show respective uncertainties in the parallax values. 
	 The red circles represent the probable cluster members (P$_\mu>$ 70\%). The green squares represent the probable cluster members with 
	 good parallax estimates ($\varpi$/$\sigma$ $_\varpi$ $>$ 5).}
\label{fig4}
\end{figure}

A circular area of radius 0.7 mas yr$^{-1}$ around the peak of the over-density in the VPD is used to select the probable cluster members, and the remaining sources in the VPD are considered as field stars. 
To determine the membership probability of the stars in the cluster region, we use the approach discussed in \citet{Pandey2020}. 
Assuming a distance of $\sim$ 1.14 kpc (see Sec. 3.4) and a radial velocity dispersion of 1 km s$^{-1}$ for open clusters \citep{Girard1989}, 
a dispersion ($\sigma_{c}$) of $\sim$ 0.19 mas yr$^{-1}$ in the PMs of the cluster can be obtained. We obtain $\mu$$_{xf}$ = -0.08 mas yr$^{-1}$, 
$\mu$$_{yf}$ = -0.33 mas yr$^{-1}$, $\sigma$$_{xf}$ = 1.05 mas yr$^{-1}$ and $\sigma$$_{yf}$ = 2.02 mas yr$^{-1}$ for the probable field members. These values are further used to construct 
the frequency distributions of the cluster stars (${\phi }_{c}^{\nu }$) and field stars (${\phi }_{f}^{\nu }$) by using the equation given in \citet{Yadav2013} 
and then the value of membership probability for the i$^{th}$ star is calculated using the equation given below:
\begin{equation}
P_\mu(i) = \frac{{n_c\times\phi^\nu_c(i)}}{n_c\times\phi^\nu_c(i)+n_f\times\phi^\nu_f(i)}
\end{equation}
where n$_c$ and n$_f$ are the normalized number of probable cluster members and field members, respectively. 
In Figure~\ref{fig4} (upper panel), we have plotted the estimated membership probability for all the Gaia sources within the cluster radius as a function of G-band magnitude as blue dots.
Gaia sources with high membership probability (P$_\mu>$ 70\%) are shown with red circles. 217 stars has been found with (P$_\mu>$ 70\%). There is a clear separation between the cluster members and field stars toward the brighter part, supporting the effectiveness of this technique. A high membership probability extends down to G $\sim$ 19 mag, whereas toward the fainter limits, the probability gradually decreases. 
A majority of the stars with high membership probability follow a tight distribution in the VPD.

We have also plotted the parallax for all the Gaia sources as a function of G-band magnitude (dots in Figure \ref{fig4}, bottom panel). The respective uncertainties in the parallax values are shown with the error bars. The red circles represent sources having membership probability 
P$_\mu>$ 70\%. We estimate the cluster distance using the parallax values of the cluster members having high membership probability (P$_\mu>$ 70\%) and 
good parallax accuracy ($\varpi$/$\sigma_\varpi$ $>$5). These sources are shown with the red circles in Figure \ref{fig4} (bottom panel). The median parallax value of these sources 
is 0.892 $\pm$ 0.088 mas. We estimate the cluster distance after correcting the median parallax value for the known parallax offset of $\sim$ -0.015 \citep{Stassun2021}. 
The distance estimate for the cluster using Gaia data comes out to be $\sim$ 1.14 $\pm$ 0.11 kpc, which is in agreement with that reported by \citep{CantatGaudin2020}).  
Using a matching radius of $\sim$ 1 arcsec, we get $\sim$ 1610 optical $UBVR_cI_c$ counterparts of Gaia sources with membership probability information within the cluster radius. Out of these, about 160 stars are cluster members. 

\subsection{Reddening toward the NGC 1027: (U-B)/(B-V) Color-Color Diagram}

\begin{figure*}
\centering
	\includegraphics[scale = 0.48, trim = 0 0 0 0, clip]{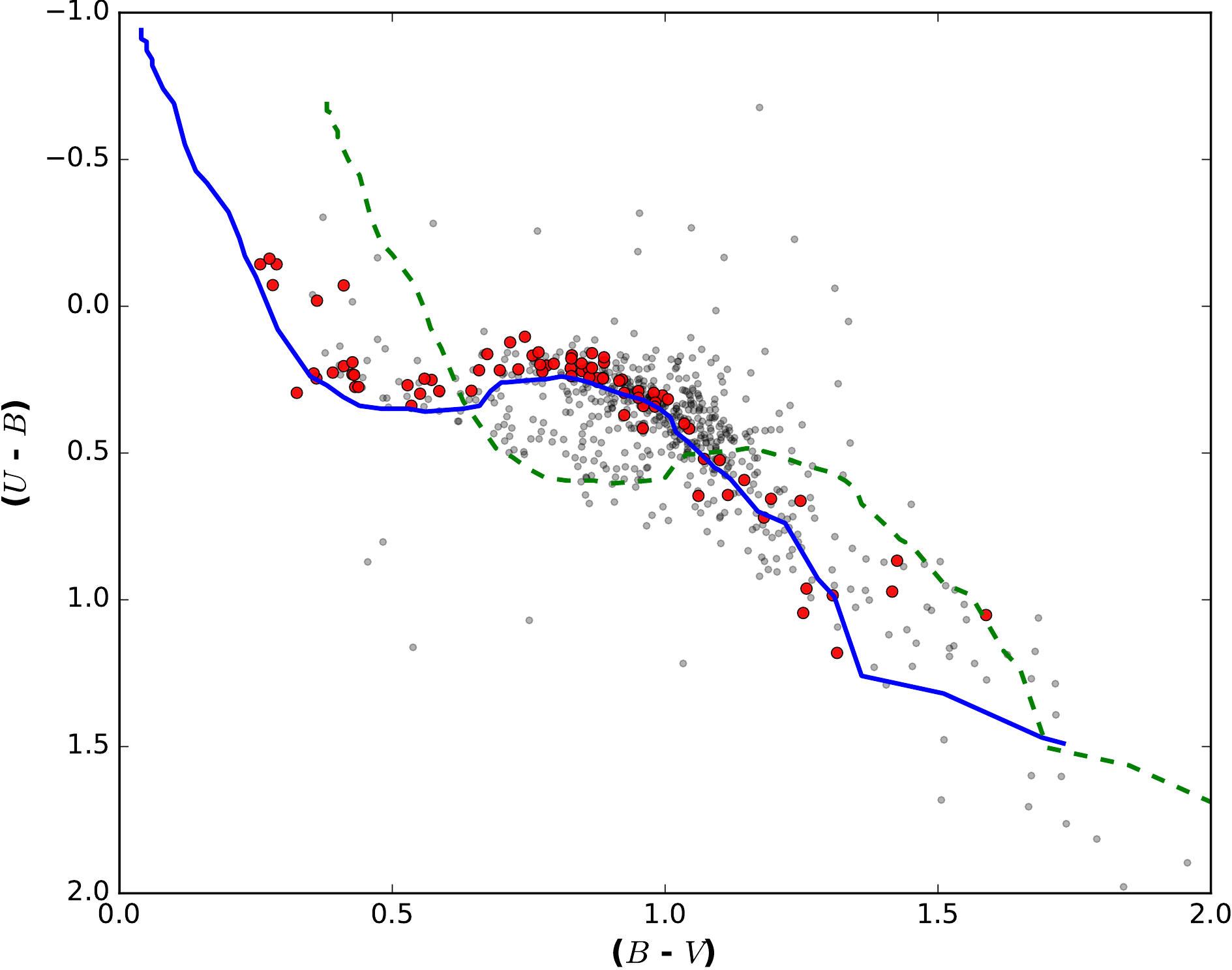}
	\caption{Two color diagram for NGC 1027. Red color circle denotes member stars present within 8 arcmin radius}
        \label{fig5}
\end{figure*}

Knowledge of reddening is essential for estimation of distance and age of the cluster. The reddening, E($B$-$V$) in the cluster region is determined using the two color diagram (TCD) ($U$-$B$) versus ($B$-$V$) for the member stars within 8 arcmin (Figure \ref{fig5}). To avoid large uncertainties we have skipped stars having photometric error larger than 0.1 mag in $UBV$ band. We adopt the slope of the reddening vector E($U$-$B$)/E($B$-$V$) = 0.72 and fit the intrinsic zero-age-main sequence (ZAMS) with the solar metallicity \citep{SchmidtKale1982} for main-sequence stars by shifting the ZAMS for different color excesses. 
The best fit corresponding to E($B$-$V$)=0.36 mag is shown with the thick curve in Figure \ref{fig5}. The dashed curve shows the ZAMS shifted for a reddening E($B$-$V$) of 0.7 mag. Our estimated value of color excess E($B-V$) = 0.36 $\pm$ 0.04 mag is comparable to E($B-V$) = 0.41$\pm$ 0.12 mag estimated by \citet{Maciejewski2007}. Line-of-sight  extinction, A$_G$, and  the  reddening,  E($G_{BP}-G_{RP}$),  are  estimated using the relations given in \citep{wang2019}. 
We use $A_G$ = 1.21 and E($G_{BP}-G_{RP}$) = 0.35 for the further analyses of Gaia EDR3 data.\\

\subsection{Age and distance of the cluster}
\begin{figure*}
\centering
	\includegraphics[scale = 0.5, trim = 0 0 0 0, clip]{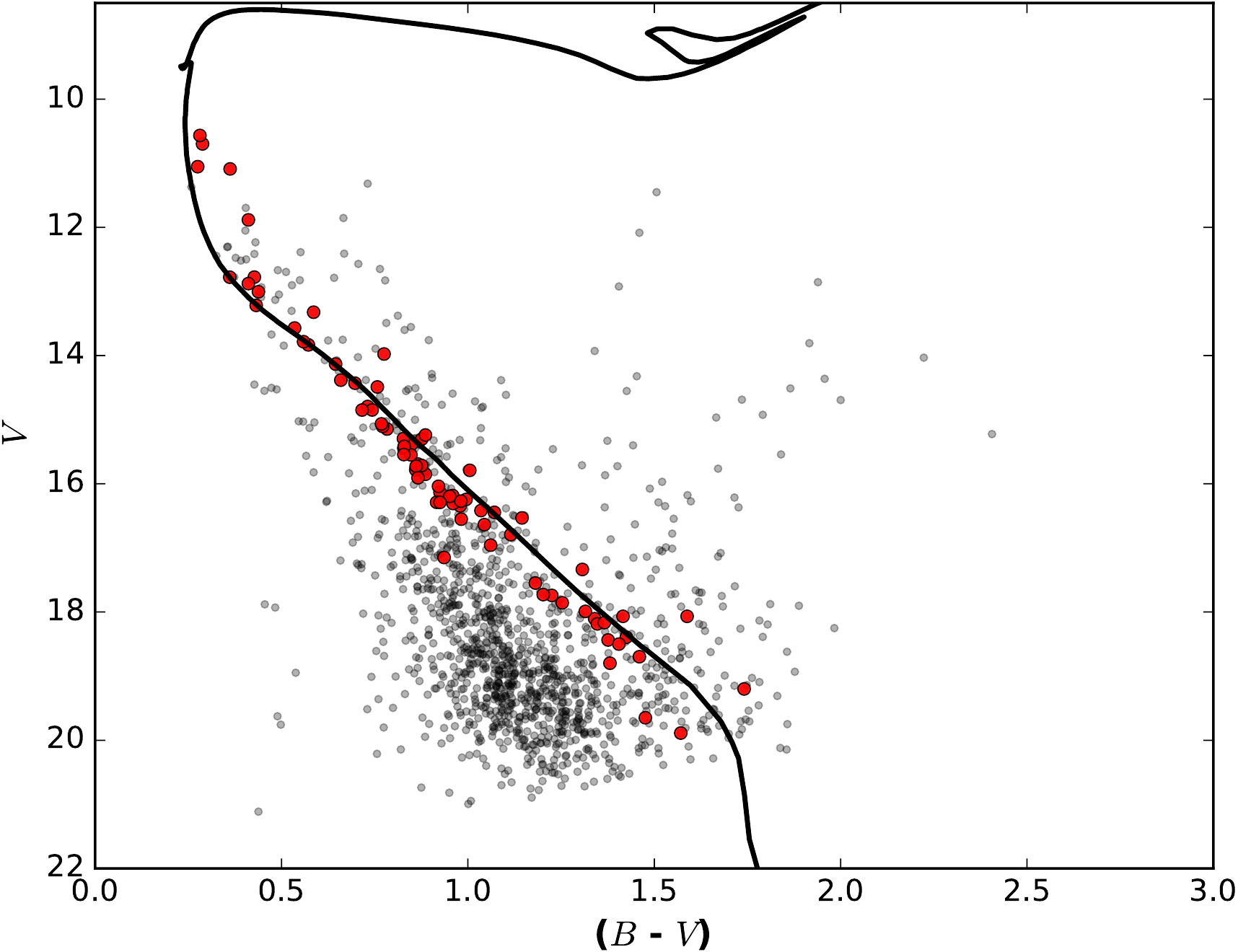}
	\includegraphics[scale = 0.5, trim = 0 0 0 0, clip]{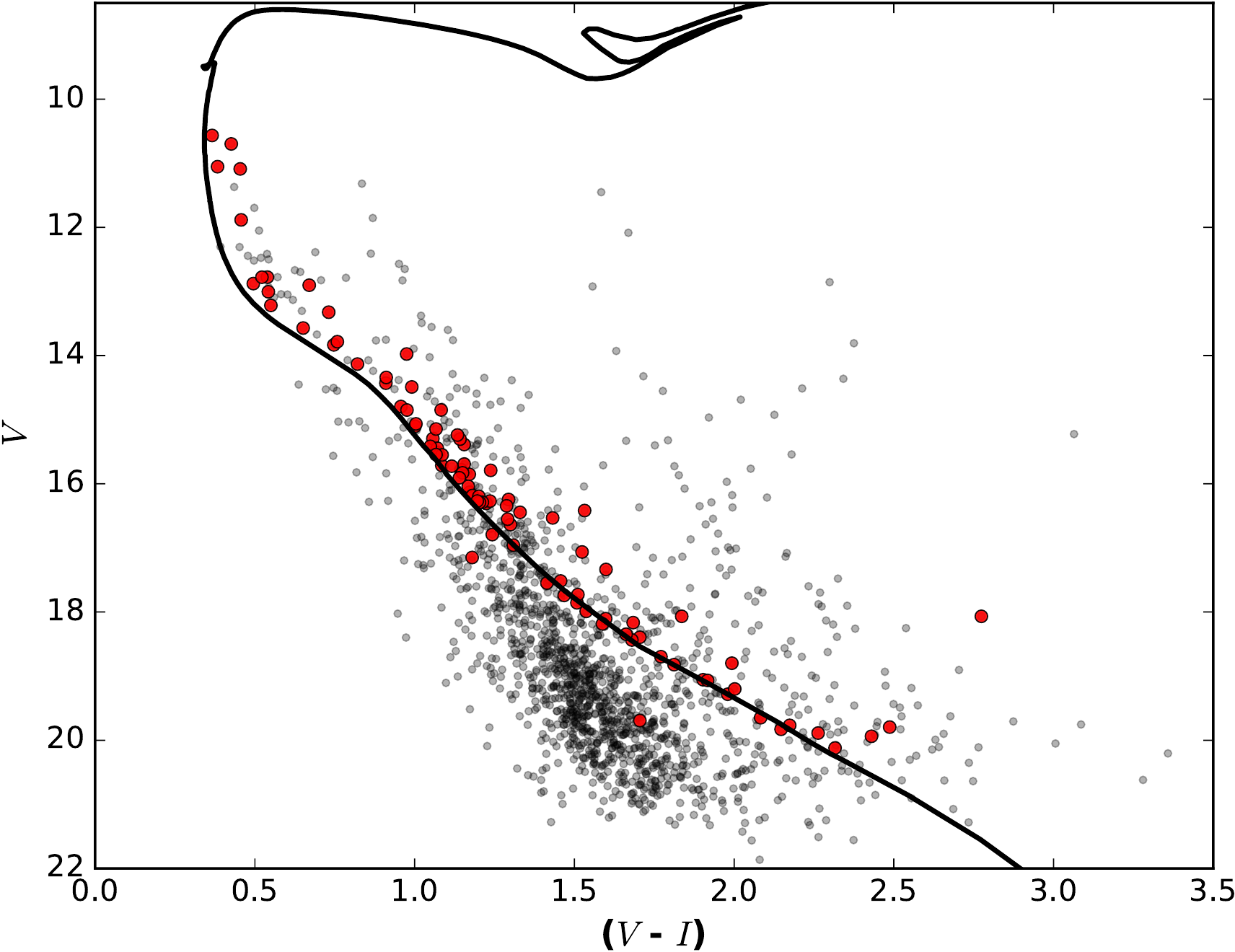}
	\caption{$V$/$(B-V)$ (left panel) and $V$/($V$-$I$) (right panel) color magnitude diagrams for the stars in the cluster NGC 1027 (radius $\le$ 8 arcmin). The red circles represent high membership probability members from Gaia EDR3.}
        \label{fig6}
\end{figure*}

\begin{figure}
\centering
\includegraphics[scale = 0.48, trim = 0 0 0 0,clip]{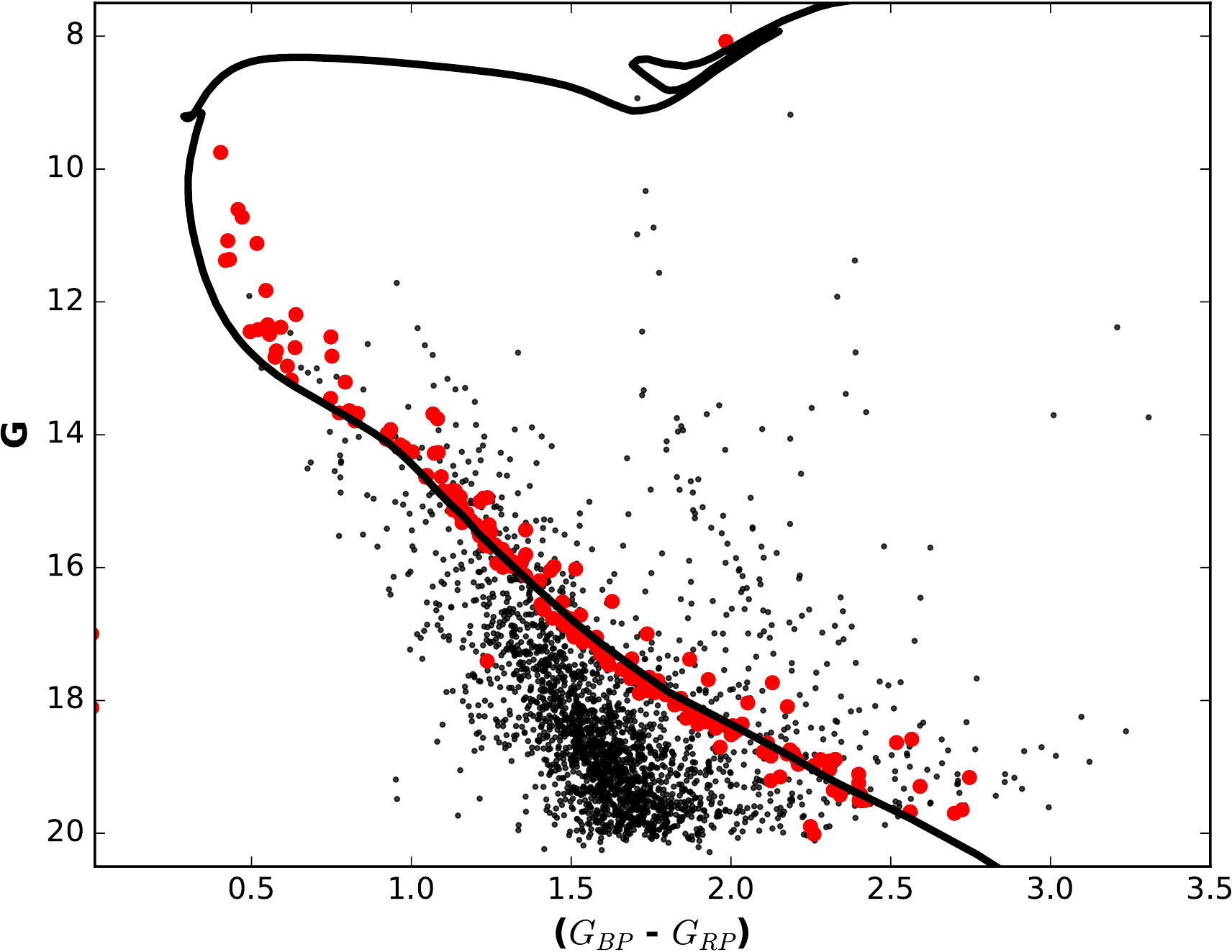}
\caption{$G$/($G_{BP}$-$G_{RP}$) color-magnitude diagram for the stars in the cluster region. The red circles represent Gaia sources with P$_\mu$ $>$ 70. The thick curve represent the main sequence isochrone (Padova-Parsec) for log age of 8.11 corrected for the adopted distance and reddening.}
\label{fig7}
\end{figure}

  The CMDs are important tools to estimate the cluster's age and distance.  We have constructed the optical CMDs using cluster members from our optical catalogue and Gaia data. $V$ vs ($B-V$) and $V$ vs ($V-I$) color-magnitude diagrams for the sources in the cluster region (black dots) along with the member stars (red circles) shown in the Figure \ref{fig6}. $G_{BP}-G_{RP}$ \ref{fig7} is plotted using the cluster members with membership probabilities P$_\mu$ $>$ 70\%.
  The CMDs of target cluster show a well-defined narrow main sequence (MS) that extends from $V\sim$10.5 mag to $V\sim$20 mag. It is seen that the field star contamination becomes more evident for stars $V$ $\ge$ 17 mag. 
  
 Distance and age estimation of the cluster have been derived by using MS isochrone for solar metallicity taken from Padova  PARSEC  version  1.2S  database  of  stellar evolutionary  tracks \&  isochrones  of  \citep{Pastorelli2020}. This is corrected for the derived reddening values from the TCD. We tried to fit the MS isochrones of different ages over the member stars (shown with the red circles in Figure \ref{fig6} and \ref{fig7}). The visual best fit of the MS to the blue part of the intrinsic CMDs gives distance modulus ($V$-M$_V$). The best fit isochrone yield a distance modulus ($V$-M$_V$) = 10.28 mag indicating a distance of $\sim$ 1.14 Kpc for the cluster and an log age of 8.11 ($\sim$ 130 Myr) corrected for the adopted distance and reddening.  Our estimated distance for the target field is comparable to the value of 1.03 Kpc whereas the age estimation is lower than the value 252 Myr derived by \citep{Maciejewski2007}.

\subsection{Luminosity Function and Mass function}
 
\begin{figure}
\includegraphics[scale = 0.48, trim = 0 0 0 0, clip]{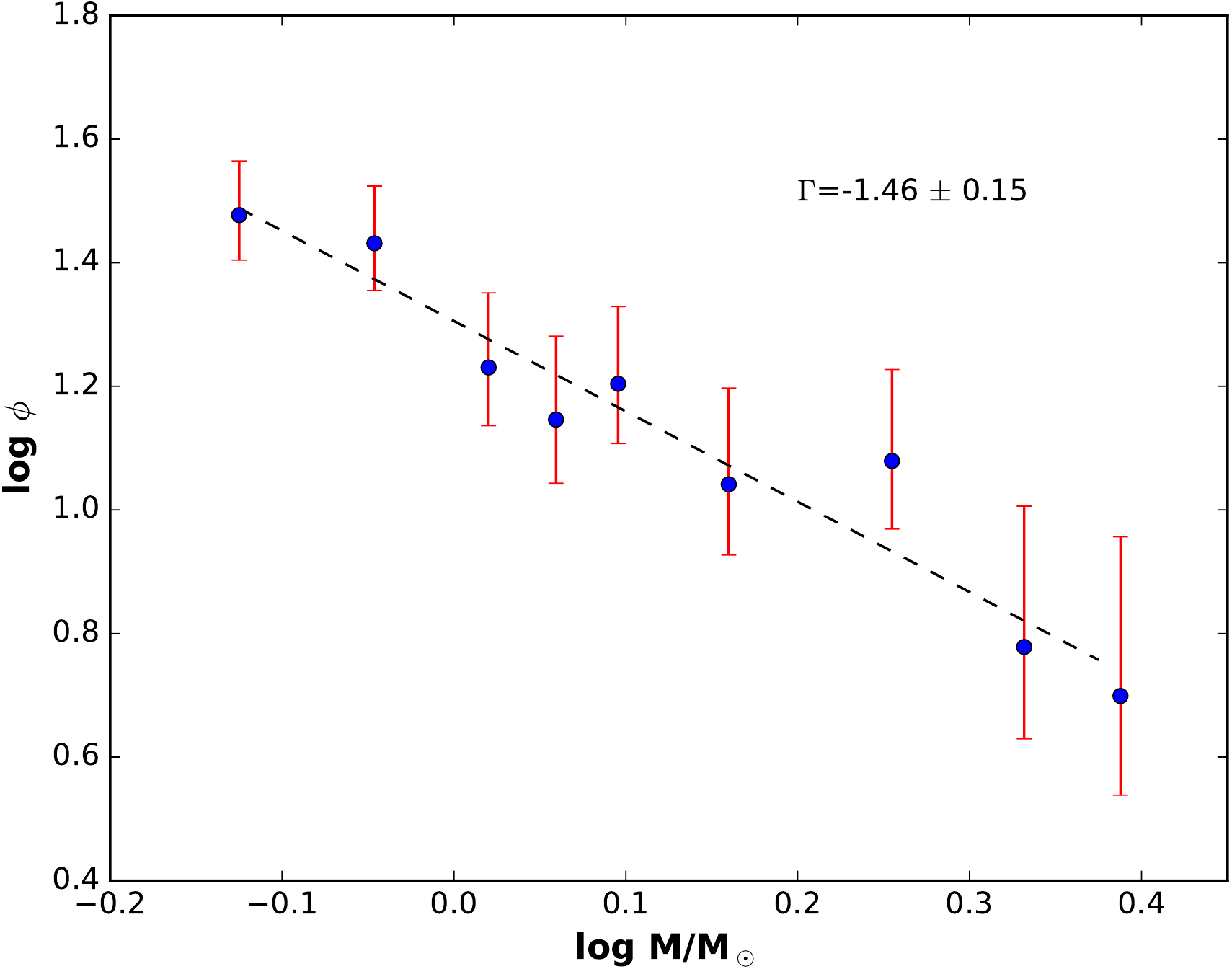}
\caption{Mass functions of stars using probable member main-sequence stars. The error bars represent 1/$\sqrt{N}$}
\label{fig8}
\end{figure}

Stellar mass cannot be measured directly. Stellar models are required to transform magnitude of a star in a given wavelength into masses. With the help of CMDs, we derive luminosity function (LF) of probable main sequence cluster members and MF using mass-luminosity relation from theoretical evolutionary models. We have used Gaia EDR3 high membership probability members to determine LFs for the cluster. We have used $G$ versus ($G_{B_P} - G_{R_P}$) CMD to construct the LF. Field star contamination and membership uncertainty affects luminosity and mass function of open star clusters \citep{Scalo1998}, therefore, to avoid the contribution of the
field stars in our sample of cluster members, we have used only kinematic and photometric confirmed members of the cluster for the LF and MF estimation. To determine the completeness of the data, we constructed the magnitude histogram method \citep{Panwar2017}. Using the magnitude histogram, we found that completeness of the data limits to $G$
$\sim$ 18.5 mag. For estimation of LF, magnitude of the MS stars were converted to absolute magnitudes using estimated
distance modulus and reddening. Conversion of  LF into present day MF is done using the latest
version of the theoretical isochrones of Padova’s stellar
evolutionary tracks and isochrones \citep{Pastorelli2020}. To
construct MF, we have converted absolute mag bins to mass bins.
Consequently, we transform the LF to MF for all the
main-sequence stars ($\sim$ 180 stars) having mass 0.55 $\le$
M/M$_\odot$ $\le$ 2.1. 
MF slope is given by using a power-law 
\begin{equation}
\Gamma = d \log N (\log m)/d \log m 
\end{equation}
where $N(\log m)$ is the number of stars per unit logarithmic
mass interval.  dN is the probable cluster members in a mass bin
dM with central mass M. Observational results suggest that the MF slopes is similar to that given by
\citet{Salpeter1955}, that is, $\Gamma$ = -1.35. The resulting mass function for our
target cluster is shown in Figure \ref{fig8}. For the cluster NGC 1027, we obtain that the mass distribution
has a best-fitting slope $\Gamma$ = -1.46 $\pm$ 0.15, similar to the
Salpeter value.

\begin{figure*}
\centering
 \includegraphics[scale = 0.45, trim = 0 0 0 0, clip]{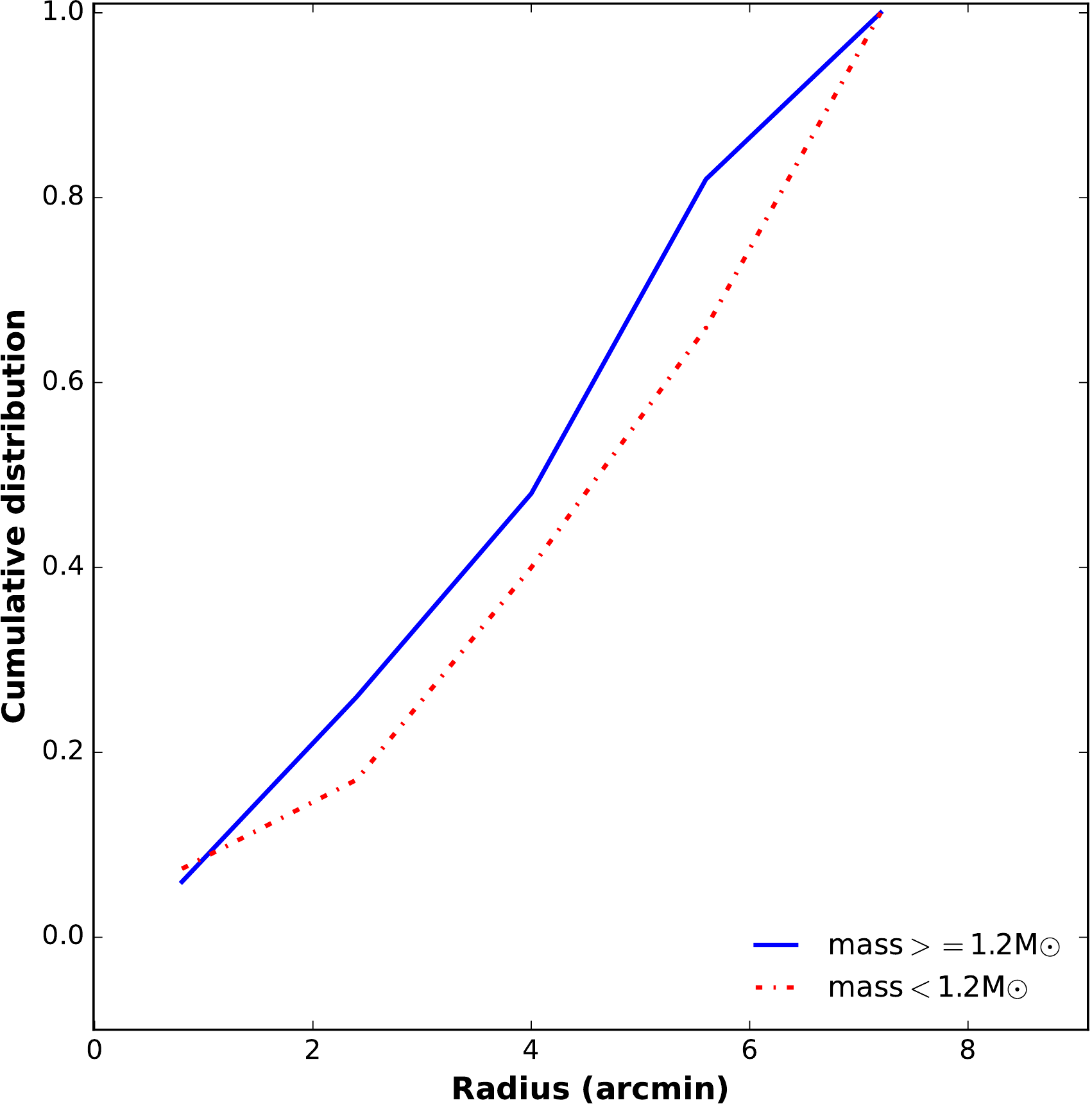}
 \includegraphics[scale = 0.5, trim = 0 0 0 0, clip]{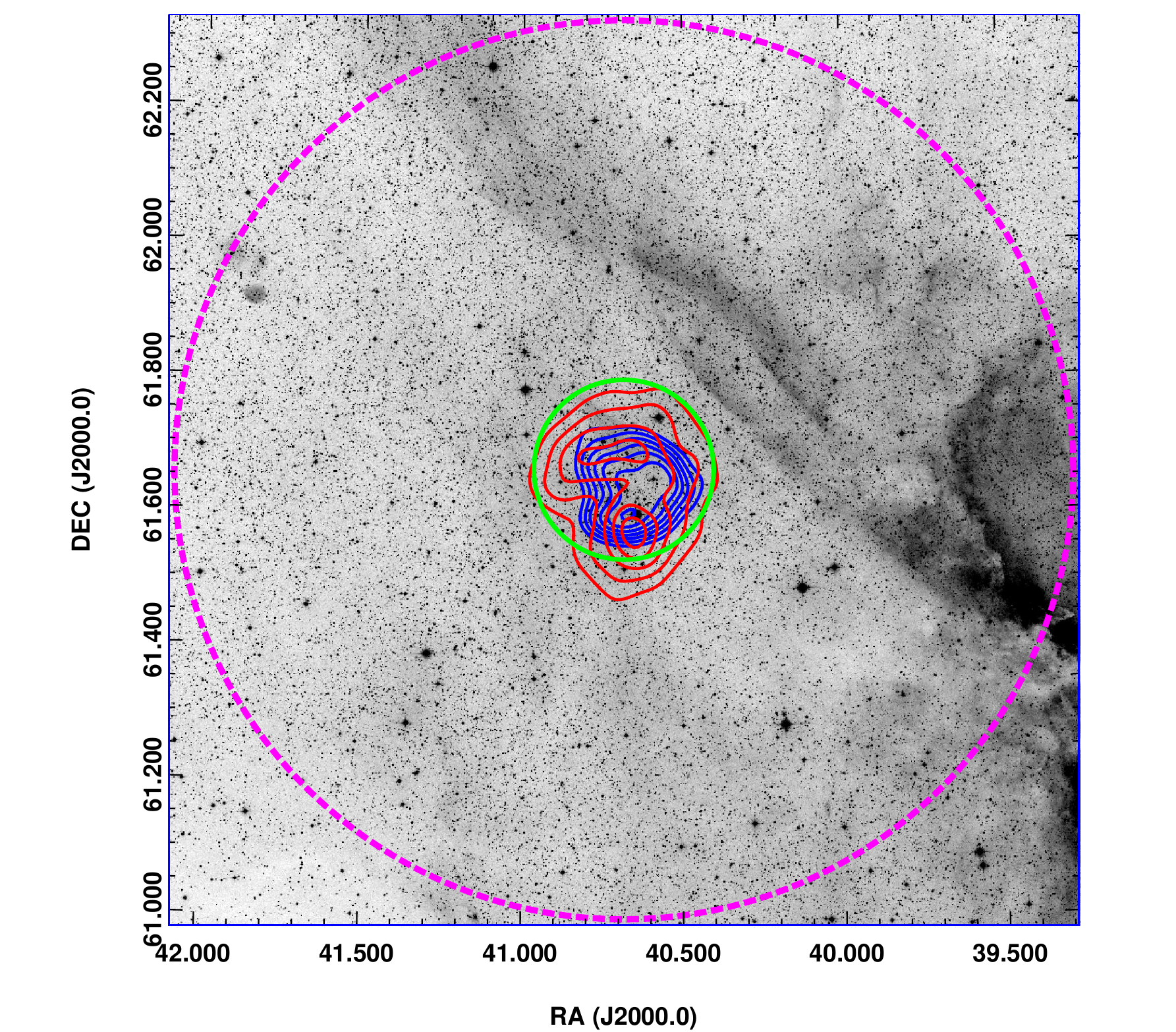}
	\caption{Left panel: Normalized cumulative radial distribution of the member stars in the cluster region. Right panel: {\bf DSS2-R image of the open cluster NGC~1027.  The red and blue contours show the density distribution of low-mass ($<$ 1.2
	M$_\odot$) and relatively higher-mass ($\ge$ 1.2 M$_\odot$) members, respectively. The green circle shows the cluster radius
	and magenta dashed circle represents the area (radius $\sim$ 40 arcmin), used for the search of kinematic members outside cluster radius. }}
    \label{fig9}
    \end{figure*}

\section{Discussion}

Observational investigation of the OCs reveal that in general, massive stars are concentrated toward the nucleus in comparison to low-mass stars which are mostly distributed in the corona region of the cluster \citep{Fischer1998,Sharma2000}.

This segregation of massive stars towards cluster's center can happen due to the equipartition of energy (dynamical evolution) or due to the star formation processes itself.
 
We have  analysed the radial distribution of probable member stars (with membership probability $>$ 70 \%)  in two mass bins,  i.e., stars with masses $<$1.2 M$_\odot$ and $\ge$1.2 M$_\odot$, to look for mass segregation effect in NGC 1027 (cf. Figure \ref{fig9} (left panel)).
{\bf The distribution clearly demonstrates that stars in the mass bin with massive stars ( $\ge$1.2 M$_\odot$) are concentrated towards the cluster center compared to low mass stars (mass bin $<$ 1.2 M$_\odot$), suggesting mass segregation effect in the NGC 1027 cluster.} To investigate whether the mass segregation in this cluster is due to star formation processes or due to dynamical relaxation, we estimated the dynamical relaxation time ($T_E$), i.e., time in which the individual
star exchange sufficient energy to achieve the velocity distribution of a Maxwellian equilibrium.
We have used the method given by \citet{Bin1987} to estimate the dynamical relaxation time ($T_E$), i.e., 

\begin{equation}
T_E = \frac{N}{8logN }\times T_{cross}
\end{equation}

where, $T_{cross}$ = D/$\sigma_V$ is the crossing time, N is the total number of stars in the cluster region of diameter $D$, and $\sigma_V$ is the velocity dispersion, with a typical value of 3 km s$^{-1}$ \citep{Sharma2020,Bisht2017}. Using the above equation, we estimated the value of
 $T_E$ as $\sim$10 Myr for the NGC~1027 cluster.
Comparing  the cluster age ($\sim$130 Myr) with its dynamical relaxation time ($\sim$10 Myr), we found that the cluster is dynamically relaxed and  the observed mass segregation is due to the result of dynamical evolution. 

Due to the mass segregation process, low-mass members may become more vulnerable for evaporation and hence ejected out of the cluster more easily [e.g., see\citep{Mathieu1984}]. This stellar evaporation results in a continuous decrease of the total mass of the cluster, and hence the gravitational binding energy of the cluster. However, external disturbances, e.g., the tidal force from nearby giant molecular clouds or star clusters, passages through Galactic
spiral arms or disks, or a shear force by Galactic differential rotation, increase the stellar evaporation of the cluster.
For a cluster, the tidal radius is the defined as a distance at which two gravitational forces; one towards the Galactic center and another towards the cluster center; balances.  Thus, tidal radius is estimated to separate gravitationally bound and unbound stars to a cluster. The tidal radius of a star cluster in the solar neighborhood is given by the equation: 
\begin{equation}
r_t =\left[{ \frac{GM_C}{2(A - B)^2}}\right]^{1/3}
\end{equation}
where, $G$ is the gravitational constant, M$_C$ is the total mass of the cluster, and A, B are the Oort constants where A = 15.3 $\pm$ 0.4 km s$^{-1}$ kpc$^{-1}$, B = -11.9 $\pm$ 0.4 km s$^{-1}$ kpc$^{-1}$ \citep{Bovy2017}. For the total cloud mass of $\sim$164 M$_\odot$ estimated from the MF distribution, the corresponding tidal radius comes out to be $\sim$7.8 pc. If we assume that $\sim$ 50\% of the cluster mass is missed due to data incompleteness, the tidal radius will be $\sim$9 pc. Therefore, the tidal radius of the cluster is much larger than the present estimate of the cluster radius (R$_{cl}$ $\sim$2.65 pc), or in other words, this cluster is already started to loose its low-mass members due to dynamics.

Thus, there is a possibility of finding member stars beyond the observed cluster boundary. For this,  we searched for the stars sharing the similar proper motion and distance as cluster members, outside the observed cluster boundary.
 
We estimated the membership probability and found $\sim$460 stars with membership probability $>$ 70\% upto a radius of 40 arcmin (cf. Figure \ref{fig9}, right panel).
 
To study the cluster dynamics and mass segregation, we divided these stars into two mass bins, stars having mass less than 1.2M$_\odot$ and stars having mass greater than 1.2M$_\odot$, and created the respective stellar density distribution maps. The stellar density distributions for above two mass bins are over-plotted on a  DSS2-R band image shown in the Figure \ref{fig9} (right panel).

There is a clear evidence of mass segregation as the stars having mass greater than 1.2M$_\odot$ are concentrated toward the center compared to the relatively low-mass stars.   

The spatial distribution of the low mass members in the cluster NGC~1027 show slightly tailed structure (cf. Figure \ref{fig9}, right panel). This might have caused due to the tidal striping of member stars from the cluster. Tidal striping of stars can happen in a cluster due to the tidal forces from a nearby massive object, disk crossing, or differential rotation of the Galaxy. Thus, the tidal stripping depends on the mass and age of a cluster. The ratio of the number of cluster inside the tidal radius to those outside, can tell us about the amount of tidal stripping already done in a cluster. For example, for Blanco 1 (age $\sim$ 100 Myr) cluster, it is found to be $\sim$3  \citep{Zhang2020}, whereas, for Coma Berenicesa cluster (age $\sim$ 700 Myr) this ratio is estimated as 0.6 \citep{Tang2019}. However, in case of an old massive star cluster Hyades (age $\sim$ 800 Myr), this ratio is estimated as $\sim$1.1 \citep{Roser2019}. NGC 1027 is relatively young and appears to be more dynamically bound. The observed radius of the cluster suggest that it is a dynamically relaxed cluster that show mass-segregation and has already lost some of its member stars and currently it is in the process of dissolution due to the influence of external tidal interactions.

\section{Summary and conclusion}
 We present the  photmetric and kinematic study of the intermediate age open cluster NGC 1027 using $UBVRI_c$ data taken with 1.04m Sampunanand telescope and Gaia EDR3 data. 
The results of the study are summarize as follows:
\begin{enumerate}
      \item Using two-dimensional KDEs analysis of the Gaia EDR3 data, we have obtained the cluster center that is about 2 arcmin away from the center mentioned in $WEBDA$. The cluster extent is estimated to be about 8.0 arcmin.
     \item The distribution of proper motion of the stars in the cluster region shows a clear separation between the cluster members and field stars. The mean proper motion for the cluster has been estimated to be ($\mu_\alpha^\star$ ,  $\mu_{\delta}$ )= (- 0.84, 2.04) mas yr$^{-1}$.
     \item Based on the membership analysis, 217 member stars are identified (P$_\mu >$ 70\%) in the cluster region. Their mean parallax value is 0.892 $\pm$ 0.088 mas corresponding to a distance of $\sim$ 1.14 kpc.
     \item The ($U$-$B$)/($B$-$V$) color-color diagram for the cluster members give a reddening E($B$-$V$) $\sim$ 0.36 $\pm$ 0.04 mag towards the cluster.
     \item The CMDs for the cluster members, constructed using UBVRI photometric data and Gaia EDR3 data, show that NGC~1027 is an intermediate age OC with an age $\sim$ 130 Myr and located at a distance of $\sim$ 1.14 kpc.
     \item The LFs and MFs are determined by considering the members selected from Gaia proper motion database and photometry. The mass function slope for the cluster, $\Gamma$ =-1.46 $\pm$ 0.15, is in good agreement with that given by Salpeter (1955) for the Galactic OCs.  
    \item We have studied the radial distribution of stellar masses within the cluster region and found that relatively  massive stars are dominantly distributed in the inner regions of the cluster suggesting mass segregation.
\item We report presence of astrometric members of the target field outside the clusters radius. The observed small size of the cluster as compare to its tidal radius, indicates mass-segregation, slightly elongated morphology and a comparison of dynamical relaxation time with the age of cluster. We conclude that the cluster is under the influence of external tidal interactions.

\end{enumerate}

 \section*{Acknowledgement}
We are thankful to anonymous referee for providing the comments that significantly improve the manuscript. NP acknowledges the financial support received through the SERB CRG/2021/005876 grant. This research has made use of the WEBDA database, operated at the Department of Theoretical Physics and Astrophysics of the Masaryk University. This work has made use of data from the European Space Agency (ESA) mission {\it Gaia} https://www.cosmos.esa.int/gaia), processed by the {\it Gaia}. Data Processing and Analysis Consortium. AT also like to acknowledge Prof. U S Pandey for his scientific suggestions and corrections. 

\bibliographystyle{apj}

\begin{thebibliography}{49 }


\bibitem[{Angelo {$et~al$.}(2019)Angelo, Piatti, Dias, \& Maia}]{Angelo2019}
Angelo, M.~S., Piatti, A.~E., Dias, W.~S., \& Maia, F. F.~S. 2019, MNRAS, 488,
  1635

\bibitem[{Bastian {$et~al$.}(2010)Bastian, Covey, \& Meyer}]{Bastian2010}
Bastian, N., Covey, K.~R., \& Meyer, M.~R. 2010, arXiv preprint arXiv:1001.2965

\bibitem[{Bate {$et~al$.}(2003)Bate, Bonnell, \& Bromm}]{Bate2003}
Bate, M.~R., Bonnell, I.~A., \& Bromm, V. 2003, MNRAS, 339, 577

\bibitem[{{Binney} \& {Tremaine}(1987)}]{Bin1987}
{Binney}, J., \& {Tremaine}, S. 1987

\bibitem[{Bisht {$et~al$.}(2017)Bisht, Yadav, \& Durgapal}]{Bisht2017}
Bisht, D., Yadav, R., \& Durgapal, A. 2017, New Astronomy, 52, 55

\bibitem[{Bisht {$et~al$.}(2019)Bisht, Yadav, Ganesh, Durgapal, Rangwal, \&
  Fynbo}]{Bisht2019}
Bisht, D., Yadav, R., Ganesh, S., {$et~al$.} 2019, MNRAS, 482, 1471

\bibitem[{{Bostanc{\i}} {$et~al$.}(2018){Bostanc{\i}}, {Yontan}, {Bilir}, {Ak},
  {G{\"u}ver}, {Ak}, {Paunzen}, {Ba{\c{s}}aran}, {Vurgun}, {Akti},
  {{\c{C}}elebi}, \& {{\"U}rg{\"u}p}}]{Bostanc2018}
{Bostanc{\i}}, Z.~F., {Yontan}, T., {Bilir}, S., {$et~al$.} 2018, Astrophysics
  and Space Science, 363, 143

\bibitem[{{Bovy}(2017)}]{Bovy2017}
{Bovy}, J. 2017, mnras, 468, L63

\bibitem[{{Cantat-Gaudin} \& {Anders}(2019)}]{CantatGaudin2019}
{Cantat-Gaudin}, T., \& {Anders}, F. 2019, VizieR Online Data Catalog,
  J/A+A/633/A99

\bibitem[{Cantat-Gaudin \& {Anders}(2020)}]{CantatGaudin2020}
Cantat-Gaudin, T., \& {Anders}, F. 2020, A$\&$A, 633, A99

\bibitem[{{Cantat-Gaudin, T.} {$et~al$.}(2018){Cantat-Gaudin, T.}, {Jordi, C.},
  {Vallenari, A.}, {Bragaglia, A.}, {Balaguer-N\'u\~nez, L.}, {Soubiran, C.},
  {Bossini, D.}, {Moitinho, A.}, {Castro-Ginard, A.}, {Krone-Martins, A.},
  {Casamiquela, L.}, {Sordo, R.}, \& {Carrera, R.}}]{CantatGaudin2018}
{Cantat-Gaudin, T.}, {Jordi, C.}, {Vallenari, A.}, {$et~al$.} 2018, A$\&$A,
  618, A93

\bibitem[{{Carraro} \& {Costa}(2007)}]{Carraro2007a}
{Carraro}, G., \& {Costa}, E. 2007, A$\&$A, 464, 573

\bibitem[{Carraro {$et~al$.}(2008)Carraro, V{\'a}zquez, \&
  Moitinho}]{Carraro2008}
Carraro, G., V{\'a}zquez, R.~A., \& Moitinho, A. 2008, A$\&$A, 482, 777

\bibitem[{{Castro-Ginard} {$et~al$.}(2018){Castro-Ginard}, {Jordi}, {Luri},
  {Julbe}, {Morvan}, {Balaguer-N{\'u}{\~n}ez}, \&
  {Cantat-Gaudin}}]{CastroGinard2018}
{Castro-Ginard}, A., {Jordi}, C., {Luri}, X., {$et~al$.} 2018, A$\&$A, 618, A59

\bibitem[{Castro-Ginard {$et~al$.}(2020)Castro-Ginard, {Jordi}, {Luri},
  {{\'A}lvarez Cid-Fuentes}, {Casamiquela}, {Anders}, {Cantat-Gaudin},
  {Mongui{\'o}}, {Balaguer-N{\'u}{\~n}ez}, {Sol{\`a}}, \&
  {Badia}}]{CastroGinard2020}
Castro-Ginard, A., {Jordi}, C., {Luri}, X., {$et~al$.} 2020, A$\&$A, 635, A45

\bibitem[{{Chauhan} {$et~al$.}(2011){Chauhan}, {Pandey}, {Ogura}, {Jose},
  {Ojha}, {Samal}, \& {Mito}}]{Chauhan2011}
{Chauhan}, N., {Pandey}, A.~K., {Ogura}, K., {$et~al$.} 2011, mnras, 415, 1202

\bibitem[{Chen {$et~al$.}(2003)Chen, Hou, \& Wang}]{Chen2003}
Chen, L., Hou, J.~L., \& Wang, J.~J. 2003, The Astronomical Journal, 125, 1397

\bibitem[{Dib {$et~al$.}(2017)Dib, Schmeja, \& Hony}]{Dib2017}
Dib, S., Schmeja, S., \& Hony, S. 2017, MNRAS, 464, 1738

\bibitem[{Elmegreen(2000)}]{Elmegreen2000}
Elmegreen, B.~G. 2000, ApJ, 530, 277

\bibitem[{{Friel}(1995)}]{Friel1995}
{Friel}, E.~D. 1995, ARA$\&$A, 33, 381

\bibitem[{{Gaia Collaboration}(2020)}]{EDR3}
{Gaia Collaboration}. 2020, VizieR Online Data Catalog, I/350

\bibitem[{{Gaia Collaboration} {$et~al$.}(2018){Gaia Collaboration}, {Helmi},
  {van Leeuwen}, {McMillan}, {Massari}, {Antoja}, {Robin}, {Lindegren},
  {Bastian}, {Arenou}, {Babusiaux}, {Biermann}, {Breddels}, {Hobbs}, {Jordi},
  {Pancino}, {Reyl{\'e}}, {Veljanoski}, {Brown}, {Vallenari}, {Prusti}, {de
  Bruijne}, {Bailer-Jones}, {Evans}, {Eyer}, {Jansen}, {Klioner}, {Lammers},
  {Luri}, {Mignard}, {Panem}, {Pourbaix}, {Randich}, {Sartoretti}, {Siddiqui},
  {Soubiran}, {Walton}, {Cropper}, {Drimmel}, {Katz}, {Lattanzi}, {Bakker},
  {Cacciari}, {Casta{\~n}eda}, {Chaoul}, {Cheek}, {De Angeli}, {Fabricius},
  {Guerra}, {Holl}, {Masana}, {Messineo}, {Mowlavi}, {Nienartowicz}, {Panuzzo},
  {Portell}, {Riello}, {Seabroke}, {Tanga}, {Th{\'e}venin}, {Gracia-Abril},
  {Comoretto}, {Garcia-Reinaldos}, {Teyssier}, {Altmann}, {Andrae}, {Audard},
  {Bellas-Velidis}, {Benson}, {Berthier}, {Blomme}, {Burgess}, {Busso},
  {Carry}, {Cellino}, {Clementini}, {Clotet}, {Creevey}, {Davidson}, {De
  Ridder}, {Delchambre}, {Dell'Oro}, {Ducourant},
  {Fern{\'a}ndez-Hern{\'a}ndez}, {Fouesneau}, {Fr{\'e}mat}, {Galluccio},
  {Garc{\'\i}a-Torres}, {Gonz{\'a}lez-N{\'u}{\~n}ez}, {Gonz{\'a}lez-Vidal},
  {Gosset}, {Guy}, {Halbwachs}, {Hambly}, {Harrison}, {Hern{\'a}ndez},
  {Hestroffer}, {Hodgkin}, {Hutton}, {Jasniewicz}, {Jean-Antoine-Piccolo},
  {Jordan}, {Korn}, {Krone-Martins}, {Lanzafame}, {Lebzelter}, {L{\"o}ffler},
  {Manteiga}, {Marrese}, {Mart{\'\i}n-Fleitas}, {Moitinho}, {Mora}, {Muinonen},
  {Osinde}, {Pauwels}, {Petit}, {Recio-Blanco}, {Richards}, {Rimoldini},
  {Sarro}, {Siopis}, {Smith}, {Sozzetti}, {S{\"u}veges}, {Torra}, {van Reeven},
  {Abbas}, {Abreu Aramburu}, {Accart}, {Aerts}, {Altavilla}, {{\'A}lvarez},
  {Alvarez}, {Alves}, {Anderson}, {Andrei}, {Anglada Varela}, {Antiche},
  {Arcay}, {Astraatmadja}, {Bach}, {Baker}, {Balaguer-N{\'u}{\~n}ez}, {Balm},
  {Barache}, {Barata}, {Barbato}, {Barblan}, {Barklem}, {Barrado}, {Barros},
  {Barstow}, {Bartholom{\'e} Mu{\~n}oz}, {Bassilana}, {Becciani}, {Bellazzini},
  {Berihuete}, {Bertone}, {Bianchi}, {Bienaym{\'e}}, {Blanco-Cuaresma}, {Boch},
  {Boeche}, {Bombrun}, {Borrachero}, {Bossini}, {Bouquillon}, {Bourda},
  {Bragaglia}, {Bramante}, {Bressan}, {Brouillet}, {Br{\"u}semeister},
  {Brugaletta}, {Bucciarelli}, {Burlacu}, {Busonero}, {Butkevich}, {Buzzi},
  {Caffau}, {Cancelliere}, {Cannizzaro}, {Cantat-Gaudin}, {Carballo},
  {Carlucci}, {Carrasco}, {Casamiquela}, {Castellani}, {Castro-Ginard},
  {Charlot}, {Chemin}, {Chiavassa}, {Cocozza}, {Costigan}, {Cowell}, {Crifo},
  {Crosta}, {Crowley}, {Cuypers}, {Dafonte}, {Damerdji}, {Dapergolas}, {David},
  {David}, {de Laverny}, {De Luise}, {De March}, {de Martino}, {de Souza}, {de
  Torres}, {Debosscher}, {del Pozo}, {Delbo}, {Delgado}, {Delgado}, {Di
  Matteo}, {Diakite}, {Diener}, {Distefano}, {Dolding}, {Drazinos},
  {Dur{\'a}n}, {Edvardsson}, {Enke}, {Eriksson}, {Esquej}, {Eynard Bontemps},
  {Fabre}, {Fabrizio}, {Faigler}, {Falc{\~a}o}, {Farr{\`a}s Casas}, {Federici},
  {Fedorets}, {Fernique}, {Figueras}, {Filippi}, {Findeisen}, {Fonti},
  {Fraile}, {Fraser}, {Fr{\'e}zouls}, {Gai}, {Galleti}, {Garabato},
  {Garc{\'\i}a-Sedano}, {Garofalo}, {Garralda}, {Gavel}, {Gavras}, {Gerssen},
  {Geyer}, {Giacobbe}, {Gilmore}, {Girona}, {Giuffrida}, {Glass}, {Gomes},
  {Granvik}, {Gueguen}, {Guerrier}, {Guiraud}, {Guti{\'e}rrez-S{\'a}nchez},
  {Hofmann}, {Holland}, {Huckle}, {Hypki}, {Icardi}, {Jan{\ss}en}, {Jevardat de
  Fombelle}, {Jonker}, {Juh{\'a}sz}, {Julbe}, {Karampelas}, {Kewley}, {Klar},
  {Kochoska}, {Kohley}, {Kolenberg}, {Kontizas}, {Kontizas}, {Koposov},
  {Kordopatis}, {Kostrzewa-Rutkowska}, {Koubsky}, {Lambert}, {Lanza}, {Lasne},
  {Lavigne}, {Le Fustec}, {Le Poncin-Lafitte}, {Lebreton}, {Leccia}, {Leclerc},
  {Lecoeur-Taibi}, {Lenhardt}, {Leroux}, {Liao}, {Licata}, {Lindstr{\o}m},
  {Lister}, {Livanou}, {Lobel}, {L{\'o}pez}, {Managau}, {Mann}, {Mantelet},
  {Marchal}, {Marchant}, {Marconi}, {Marinoni}, {Marschalk{\'o}}, {Marshall},
  {Martino}, {Marton}, {Mary}, {Matijevi{\v{c}}}, {Mazeh}, {Messina},
  {Michalik}, {Millar}, {Molina}, {Molinaro}, {Moln{\'a}r}, {Montegriffo},
  {Mor}, {Morbidelli}, {Morel}, {Morris}, {Mulone}, {Muraveva}, {Musella},
  {Nelemans}, {Nicastro}, {Noval}, {O'Mullane}, {Ord{\'e}novic},
  {Ord{\'o}{\~n}ez-Blanco}, {Osborne}, {Pagani}, {Pagano}, {Pailler},
  {Palacin}, {Palaversa}, {Panahi}, {Pawlak}, {Piersimoni}, {Pineau}, {Plachy},
  {Plum}, {Poggio}, {Poujoulet}, {Pr{\v{s}}a}, {Pulone}, {Racero}, {Ragaini},
  {Rambaux}, {Ramos-Lerate}, {Regibo}, {Riclet}, {Ripepi}, {Riva}, {Rivard},
  {Rixon}, {Roegiers}, {Roelens}, {Romero-G{\'o}mez}, {Rowell}, {Royer},
  {Ruiz-Dern}, {Sadowski}, {Sagrist{\`a} Sell{\'e}s}, {Sahlmann}, {Salgado},
  {Salguero}, {Sanna}, {Santana-Ros}, {Sarasso}, {Savietto}, {Schultheis},
  {Sciacca}, {Segol}, {Segovia}, {S{\'e}gransan}, {Shih}, {Siltala}, {Silva},
  {Smart}, {Smith}, {Solano}, {Solitro}, {Sordo}, {Soria Nieto}, {Souchay},
  {Spagna}, {Spoto}, {Stampa}, {Steele}, {Steidelm{\"u}ller}, {Stephenson},
  {Stoev}, {Suess}, {Surdej}, {Szabados}, {Szegedi-Elek}, {Tapiador}, {Taris},
  {Tauran}, {Taylor}, {Teixeira}, {Terrett}, {Teyssandier}, {Thuillot},
  {Titarenko}, {Torra Clotet}, {Turon}, {Ulla}, {Utrilla}, {Uzzi}, {Vaillant},
  {Valentini}, {Valette}, {van Elteren}, {Van Hemelryck}, {van Leeuwen},
  {Vaschetto}, {Vecchiato}, {Viala}, {Vicente}, {Vogt}, {von Essen}, {Voss},
  {Votruba}, {Voutsinas}, {Walmsley}, {Weiler}, {Wertz}, {Wevems},
  {Wyrzykowski}, {Yoldas}, {{\v{Z}}erjal}, {Ziaeepour}, {Zorec}, {Zschocke},
  {Zucker}, {Zurbach}, \& {Zwitter}}]{GDR2}
{Gaia Collaboration}, {Helmi}, A., {van Leeuwen}, F., {$et~al$.} 2018, aap,
  616, A12

\bibitem[{Gao(2014)}]{Gao2014}
Gao, X.-H. 2014, Research in Astronomy and Astrophysics, 14, 159

\bibitem[{Girard {$et~al$.}(1989)Girard, Grundy, L{\'o}pez, \& van
  Altena}]{Girard1989}
Girard, T.~M., Grundy, W.~M., L{\'o}pez, C.~E., \& van Altena, W.~F. 1989, AJ,
  98, 227

\bibitem[{Harris \& Pudritz(1994)}]{Harris1994}
Harris, W.~E., \& Pudritz, R.~E. 1994, ApJ, 429, 177

\bibitem[{Joshi {$et~al$.}(2016)Joshi, Dambis, Pandey, \& Joshi}]{Joshi2016}
Joshi, Y.~C., Dambis, A., Pandey, A.~K., \& Joshi, S. 2016, A\&A, 593, A116

\bibitem[{Joshi {$et~al$.}(2020)Joshi, Maurya, John, Panchal, Joshi, \&
  Kumar}]{Joshi2020}
Joshi, Y.~C., Maurya, J., John, A.~A., {$et~al$.} 2020, MNRAS, 492, 3602

\bibitem[{Kharchenko {$et~al$.}(2013)Kharchenko, Piskunov, Schilbach,
  R{\"o}ser, \& Scholz}]{Kharchenko2013}
Kharchenko, N., Piskunov, A., Schilbach, E., R{\"o}ser, S., \& Scholz, R.-D.
  2013, A$\&$A, 558, A53

\bibitem[{{King}(1962)}]{King1962}
{King}, I. 1962, AJ, 67, 471

\bibitem[{Kroupa(2002)}]{Kroupa2002}
Kroupa, P. 2002, Science, 295, 82

\bibitem[{Lada \& Lada(2003)}]{Lada&Lada2003}
Lada, C.~J., \& Lada, E.~A. 2003, ARA$\&$A, 41, 57

\bibitem[{Landolt(1992)}]{Landolt1992}
Landolt, A.~U. 1992, AJ, 104, 340

\bibitem[{Liu \& Pang(2019)}]{Liu2019}
Liu, L., \& Pang, X. 2019, The Astrophysical Journal Supplement Series, 245, 32

\bibitem[{Maciejewski \& Niedzielski(2007)}]{Maciejewski2007}
Maciejewski, G., \& Niedzielski, A. 2007, A\&A, 467, 1065

\bibitem[{{Mathieu}(1984)}]{Mathieu1984}
{Mathieu}, R.~D. 1984, apj, 284, 643

\bibitem[{Medhi \& Tamura(2013)}]{MedhiandTamura2013}
Medhi, B.~J., \& Tamura, M. 2013, MNRAS, 430, 1334

\bibitem[{{Miholics} {$et~al$.}(2014){Miholics}, {Webb}, \&
  {Sills}}]{Miholics2014}
{Miholics}, M., {Webb}, J.~J., \& {Sills}, A. 2014, MNRAS, 445, 2872

\bibitem[{Pandey {$et~al$.}(2020)Pandey, Sharma, Panwar, Dewangan, Ojha, Bisen,
  Sinha, Ghosh, \& Pandey}]{Pandey2020}
Pandey, R., Sharma, S., Panwar, N., {$et~al$.} 2020, ApJ, 891, 81

\bibitem[{Pang {$et~al$.}(2020)Pang, Li, Tang, Pasquato, \&
  Kouwenhoven}]{Pang2020}
Pang, X., Li, Y., Tang, S.-Y., Pasquato, M., \& Kouwenhoven, M. 2020, The
  Astrophysical Journal Letters, 900, L4

\bibitem[{Panwar {$et~al$.}(2017)Panwar, Samal, Pandey, Jose, Chen, Ojha,
  Ogura, Singh, \& Yadav}]{Panwar2017}
Panwar, N., Samal, M., Pandey, A., {$et~al$.} 2017, MNRAS, 468, 2684

\bibitem[{{Pastorelli} {$et~al$.}(2020){Pastorelli}, {Marigo}, {Girardi},
  {Aringer}, {Chen}, {Rubele}, {Trabucchi}, {Bladh}, {Boyer}, {Bressan},
  {Dalcanton}, {Groenewegen}, {Lebzelter}, {Mowlavi}, {Chubb}, {Cioni}, {de
  Grijs}, {Ivanov}, {Nanni}, {van Loon}, \& {Zaggia}}]{Pastorelli2020}
{Pastorelli}, G., {Marigo}, P., {Girardi}, L., {$et~al$.} 2020, MNRAS, 498,
  3283

\bibitem[{Phelps \& Janes(1994)}]{Phelps1994}
Phelps, R.~L., \& Janes, K.~A. 1994, The Astrophysical Journal Supplement
  Series, 90, 31

\bibitem[{{R{\"o}ser} {$et~al$.}(2019){R{\"o}ser}, {Schilbach}, \&
  {Goldman}}]{Roser2019}
{R{\"o}ser}, S., {Schilbach}, E., \& {Goldman}, B. 2019, Astronomy \&
  Astrophysics, 621, L2

\bibitem[{{Ruprecht}(1966)}]{Ruprecht1966}
{Ruprecht}, J. 1966, Bulletin of the Astronomical Institute of Czechoslovakia,
  vol. 17, p.98, 98

\bibitem[{Sagar {$et~al$.}(1986)Sagar, Piskunov, Miakutin, \&
  Joshi}]{Sagar1986}
Sagar, R., Piskunov, A., Miakutin, V., \& Joshi, U. 1986, MNRAS, 220, 383

\bibitem[{Salpeter(1955)}]{Salpeter1955}
Salpeter, E.~E. 1955, ApJ, 121, 161

\bibitem[{{Scalo}(1998)}]{Scalo1998}
{Scalo}, J. 1998, The Stellar Initial Mass Function (38th Herstmonceux
  Conference), 142, 201

\bibitem[{{Schmidt-Kaler}(1982)}]{SchmidtKale1982}
{Schmidt-Kaler}, T. 1982, A$\&$A, 19

\bibitem[{Sharma {$et~al$.}(2000)Sharma, Tuzla, Matsen, \& Chen}]{Sharma2000}
Sharma, A.~K., Tuzla, K., Matsen, J., \& Chen, J.~C. 2000, Powder Technology,
  111, 114

\bibitem[{{Sharma} {$et~al$.}(2020){Sharma}, {Ghosh}, {Ojha}, {Pandey},
  {Sinha}, {Pandey}, {Ghosh}, {Panwar}, \& {Pandey}}]{Sharma2020}
{Sharma}, S., {Ghosh}, A., {Ojha}, D.~K., {$et~al$.} 2020, MNRAS, 498, 2309

\bibitem[{Soderblom {$et~al$.}(1998)Soderblom, King, Hanson, Jones, Fischer,
  Stauffer, \& Pinsonneault}]{Fischer1998}
Soderblom, D.~R., King, J.~R., Hanson, R.~B., {$et~al$.} 1998, ApJ, 504, 192

\bibitem[{Stassun \& Torres(2021)}]{Stassun2021}
Stassun, K.~G., \& Torres, G. 2021, The Astrophysical Journal Letters, 907, L33

\bibitem[{Stetson(1987)}]{Stetson1987}
Stetson, P.~B. 1987, Publications of the Astronomical Society of the Pacific,
  99, 191

\bibitem[{{Stetson}(2000)}]{Stetson2000}
{Stetson}, P.~B. 2000, Publications of the ASP, 112, 925

\bibitem[{{Tang} {$et~al$.}(2019){Tang}, {Pang}, {Yuan}, {Chen}, {Hong},
  {Goldman}, {Just}, {Shukirgaliyev}, \& {Lin}}]{Tang2019}
{Tang}, S.-Y., {Pang}, X., {Yuan}, Z., {$et~al$.} 2019, ApJ, 877, 12

\bibitem[{{Tripathi} {$et~al$.}(2013){Tripathi}, {Pandey}, \&
  {Kumar}}]{apara2013}
{Tripathi}, A., {Pandey}, U.~S., \& {Kumar}, B. 2013, Bulletin of the
  Astronomical Society of India, 41, 209

\bibitem[{{Tripathi} {$et~al$.}(2014){Tripathi}, {Pandey}, \&
  {Kumar}}]{apara2014}
{Tripathi}, A., {Pandey}, U.~S., \& {Kumar}, B. 2014, New Astronomy, 29, 1

\bibitem[{{Trumpler}(1930)}]{Trumpler1930}
{Trumpler}, R.~J. 1930, Lick Observatory Bulletin, 420, 154

\bibitem[{{Wang} \& {Chen}(2019)}]{wang2019}
{Wang}, S., \& {Chen}, X. 2019, apj, 877, 116

\bibitem[{Yadav {$et~al$.}(2013)Yadav, Sariya, \& Sagar}]{Yadav2013}
Yadav, R. K.~S., Sariya, D.~P., \& Sagar, R. 2013, MNRAS, 430, 3350

\bibitem[{{Zhang} {$et~al$.}(2020){Zhang}, {Tang}, {Chen}, {Pang}, \&
  {Liu}}]{Zhang2020}
{Zhang}, Y., {Tang}, S.-Y., {Chen}, W.~P., {Pang}, X., \& {Liu}, J.~Z. 2020,
  apj, 889, 99

\end{thebibliography}

\end{document}